\documentclass[twocolumn,showpacs,preprintnumbers]{revtex4}
\usepackage{amsmath}
\usepackage{longtable}
\usepackage{graphicx}

\begin{document}

\preprint{}
\title{Resonance analysis of $^{147}$Sm(\textit{n},$\alpha $) cross
sections: Comparison to optical model calculations and indications of
non-statistical effects}
\author{P. E. Koehler}
\email{koehlerpe@ornl.gov}
\affiliation{Physics Division, Oak Ridge National Laboratory, Oak Ridge, Tennessee 37831}
\author{Yu. M. Gledenov}
\affiliation{Joint Institute for Nuclear Research, Dubna, Russia}
\author{T. Rauscher and C. Fr\"ohlich}
\affiliation{Departement f\"{u}r Physik und Astronomie, Universit\"{a}t Basel, CH-4056
Basel, Switzerland}
\date{\today }

\begin{abstract}
We have measured the $^{147}$Sm(\textit{n},$\alpha $) cross section from 3
eV to 500 keV and performed an $\mathcal{R}$-matrix analysis in the resolved
region ($E_{n}$$<$ 700 eV) to extract $\alpha $ widths for 104 resonances.
We computed strength functions from these resonance parameters and compared
them to transmission coefficients calculated using optical model potentials
similar to those employed as inputs to statistical model calculations. The
statistical model often is used to predict cross sections and astrophysical
reaction rates. Comparing resonance parameters rather than cross sections
allows more direct tests of potentials used in the model and hence should
offer greater insight into possible improvements. In particular, an improved 
$\alpha $+nucleus potential is needed for applications in nuclear
astrophysics. In addition to providing a more direct test of the $\alpha $%
+nucleus potential, the $\alpha $-width distributions show indications of
non-statistical effects.
\end{abstract}

\maketitle

\section{\protect\bigskip Introduction}

Reactions involving $\alpha $ particles and intermediate-to-heavy mass
nuclides often can play an important role in nucleosynthesis occurring in
massive stars at high temperatures and in explosive environments such as
supernovae \cite{Ho99,Ra2002}. For example, accurate rates for many ($\gamma 
$,$\alpha $) reactions are needed for a better understanding of the
nucleosynthesis of the neutron deficient $A>$ 90 nuclides during the
so-called \textit{p} process. As important as these reactions are, there is
scant experimental information on their rates because the cross sections are
extremely small and many of the required ``target'' isotopes have very small
natural abundances (and hence are very expensive) or are radioactive; thus,
direct measurements are very difficult. However, most of the required rates
should be calculable to sufficient accuracy using the nuclear statistical
model, but these theoretical calculations are, at present, hampered by large
uncertainties in the $\alpha $+nucleus potential in the astrophysically
relevant energy range.

We recently \cite{Gl2000} have shown that (\textit{n},$\alpha $) cross
section measurements on intermediate-to-heavy mass nuclides offer perhaps
the best means for constraining the many parameters defining a realistic $%
\alpha $+nucleus potential. Although this approach is showing great promise,
one drawback is that the comparison between theory and experiment as cross
sections involves sensitivity to other model parameters in addition to the $%
\alpha $+nucleus potential. A more direct comparison of $\alpha $ strength
functions extracted from the measurements to theoretical expectations should
avoid these confounding dependencies and allow greater insight into possible
improvements in the model.

$^{147}$Sm appears to be the best candidate for such a study for several
reasons. First, the Q-value for $^{147}$Sm(\textit{n},$\alpha $) is fairly
large so the cross section, although still very small, is expected to be
among the largest in this mass range. Second, the natural abundance of $%
^{147}$Sm is fairly large so the necessary isotopically enriched sample is
affordable. Third, unlike several other potential candidates, there is
relatively complete information on the parameters for the 100 lowest-energy
resonances. This is important because reliably extracting $\alpha $ widths
requires reasonably complete knowledge of the neutron and $\gamma $ widths
for the resonances, and comparison to theory is most meaningful if the spins
and parities of the resonances are known. For \textit{s}-wave neutrons
incident on $^{147}$Sm ($I^{\pi }=\frac{7^{-}}{2}$), $J^{\pi }=3^{-}$ and $%
4^{-}$ resonances can be formed in the compound nucleus $^{148}$Sm.

The $\alpha $ widths for several $^{147}$Sm(\textit{n},$\alpha $) resonances
for $E_{n}$ $<$ 700 eV were determined in previous measurements \cite%
{Kv70,Po72,Ba76,An80,An84}\ of this cross section, and these data were
compared to statistical model predictions. However, these comparisons were
hampered by the fact that the $\alpha $ widths for most of the resonances in
this region could not be measured. We have employed an improved detector to
make a new measurement of the $^{147}$Sm(\textit{n},$\alpha $) cross
section. In addition to extending the range of the measurements to much
higher energies, this new detector resulted in a much improved
signal-to-noise ratio in the resolved resonance region so that the $\alpha $
widths for most of the resonances in this region could be extracted from the
data. This much improved $\alpha $-width information makes possible a more
meaningful comparison between theory and experiment.

\section{Experiment}

The experiment has been described elsewhere \cite{Gl2000,Ra2003}, so only
the salient features will be given here. The measurements were made at the
Oak Ridge Electron Linear Accelerator (ORELA) \cite{Pe82,Bo90,Gu97b} white
neutron source. The ORELA was operated at a repetition rate of 525 Hz, a
power of 6 -- 8 kW and a pulse width of 8 ns. A 0.76-mm thick Cd filter was
used to eliminate the overlap of slow neutrons from previous pulses and a
1.27-cm thick Pb filter was used to help reduce overload effects from the $%
\gamma $ flash at the start of each neutron pulse. Neutron energies were
measured via time-of-flight. The detector was a compensated ionization
chamber (CIC) \cite{Ko95}. Although a CIC can have poorer pulse-height
resolution than, for example, a gridded ionization chamber, it reduces
overload effects due to the $\gamma $ flash at the start of each neutron
pulse by several orders of magnitude, allowing measurements to be made with
excellent signal-to-noise ratio to much higher neutron energies (500 keV in
the present case).

The source-to-sample distance was 8.835 m and the neutron beam was
collimated to 10 cm in diameter at the sample position. Two samples were
placed back-to-back in the center of our parallel-plate CIC with the planes
of the samples perpendicular to the neutron beam. Hence, the cross section
was measured over nearly the entire 4$\pi $ solid angle. The samples were in
the form of Sm$_{2}$O$_{3}$ enriched to 95.3\% in $^{147}$Sm and were 5.0
mg/cm$^{2}$ thick by 11 cm in diameter. The $^{6}$Li(\textit{n},$\alpha
)^{3} $H reaction was used to measure the energy dependence of the flux and
to normalize the raw counts to absolute cross section. These measurements
were made during a run when one of the $^{147}$Sm samples was replaced by a $%
^{6}$Li sample. The neutron energy scale also was calibrated during this run
using dips in the time-of-flight spectrum caused by resonances in the Cd
filter and Al vacuum windows in the flight path as well as the peak due to
the 244.5-keV resonance from the $^{6}$Li(\textit{n},$\alpha $)$^{3}$H
reaction. A $^{6}$Li sample in a separate parallel-plate CIC was used as a
flux monitor. The most recent ENDF evaluation\ \cite{Ca93}\ for the $^{6}$Li(%
\textit{n},$\alpha )^{3}$H reaction was used in calculating the absolute
cross sections. The data were corrected for the small background due to the
spontaneous $\alpha $ decay of $^{147}$Sm and for the effects of $\alpha $
straggling in the samples. This latter correction (14\%) was calculated
using the computer code SRIM \cite{Zi99}. The overall normalization
uncertainty of approximately 6\% is dominated by the uncertainty ($\pm $
4\%) in this correction and by uncertainties ($\pm $ 3\%) in the sample
sizes.

\section{Resonance analysis and results}

\begin{figure}[tbp]
\includegraphics*[width=90mm,keepaspectratio] {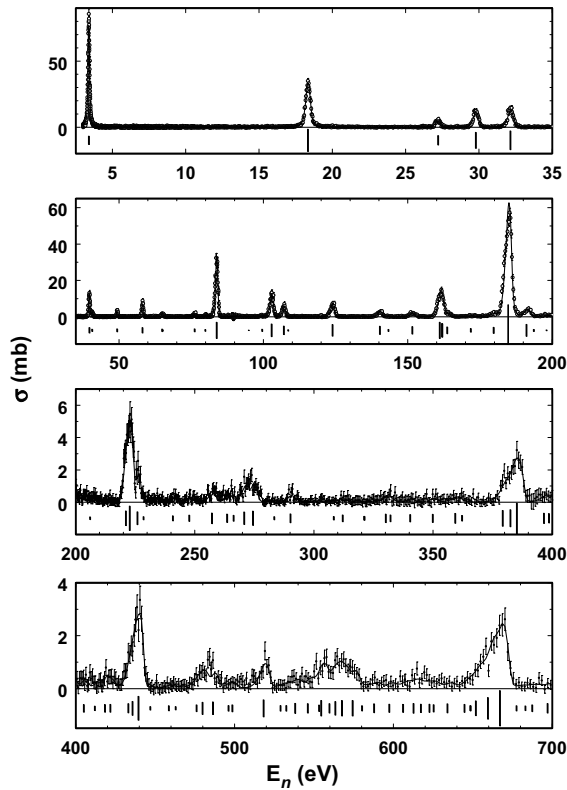}
\caption{$^{147}$Sm(\textit{n},$\protect\alpha $) cross section data (points
with error bars) and SAMMY fit (solid curves). Bars below the data show the
locations of the fitted resonances. The length of each bar is proportional
to the square root of the resonance area.}
\label{RmatrixFitsFig}
\end{figure}

The data were fitted with the $\mathcal{R}$-matrix code SAMMY \cite{La2000}
to extract the $\alpha $ widths for resonances in the resolved region below
700 eV. Almost all observed \cite{Su98} resonances in this energy range have
been assigned \cite{Ge93} as $J^{\pi }=3^{-}$ or $4^{-}$ (\textit{s}-wave).
A radius of 8.3 fm was used in all $^{147}$Sm channels. Both the fact that
the sample was an oxide and the aluminum backing were included in the input
files for SAMMY so that corrections could be applied for attenuation and
multiple-scattering effects in the sample and its backing. The resonance
energies, spins, parities, and neutron and $\gamma $ widths from the
compilation of Ref. \cite{Su98}, which are based mainly on the work of Refs. %
\cite{Co71,Ei74,Mi81,Ge93}, were used as starting values in the analysis.
Only a few of the energies had to be adjusted, and one (tentative) spin
assignment (for the 659.4-eV resonance) was changed to fit the data. A
radiation width equal to the average for $^{147}$Sm resonances (69 meV Ref. %
\cite{Mu84,Su98}) was used for resonances without $\Gamma _{\gamma }$ values
in Ref. \cite{Su98}. The resulting parameters are given in Table \ref%
{Sm147ResTable} where, for example, 1.273 \textit{(61)} is used to denote
1.273$\pm $0.061, etc. Representative plots of the data and $\mathcal{R}$%
-matrix fits are shown in Fig. \ref{RmatrixFitsFig}.

\begin{longtable}{cccccc}
\caption{$^{147}$Sm($n,\alpha$) resonance parameters. \label{Sm147ResTable}}\\%
\hline\hline
 E$_{n}$ & $J^{\pi }$ & 2$g\Gamma _{n}$ & $\mathbf{\Gamma }%
_{\gamma }$ & $\mathbf{\Gamma }_{\alpha }$ & $\frac{g\Gamma _{n}\Gamma
_{\alpha }}{\Gamma }$ \\ 
 (eV) &  & (meV) & (meV) & ($\mu $eV) & ($\mu $eV) \\ \hline
 3.397 & 3$^{-}$ & 1.18 \textit{(2)} & 67 \textit{(3)} & 1.273 \textit{(61)} & 
0.010898 \textit{(94)} \\ 
 18.340 & 4$^{-}$ & 80.9 \textit{(4)} & 72 \textit{(4)} & 0.2789 \textit{(90)}
& 0.0784 \textit{(12)} \\ 
 27.218 & 3$^{-}$ & 6.08 \textit{(11)} & 84 \textit{(5)} & 0.420 \textit{(37)} & 
0.01403 \textit{(93)} \\ 
 29.791 & 3$^{-}$ & 12.9 \textit{(2)} & 71 \textit{(6)} & 0.543 \textit{(44)} & 
0.0408 \textit{(15)} \\ 
 32.151 & 4$^{-}$ & 43.9 \textit{(6)} & 70 \textit{(5)} & 0.276 \textit{(16)}
& 0.0556 \textit{(18)} \\ 
 39.700 & 4$^{-}$ & 80.2 \textit{(11)} & 68 \textit{(4)} & 0.289 \textit{(13)}
& 0.0833 \textit{(27)} \\ 
 40.720 & 3$^{-}$ & 4.7 \textit{(2)} & 69 & 0.44 \textit{(14)} & 0.0140 \textit{(13)} \\ 
 49.358 & 4$^{-}$ & 16.5 \textit{(3)} & 75 \textit{(4)} & 0.256 \textit{(24)} & 
0.0236 \textit{(19)} \\ 
 58.130 & 3$^{-}$ & 35.9 \textit{(6)} & 77 \textit{(5)} & 0.556 \textit{(36)} & 
0.0845 \textit{(39)} \\ 
 64.96 & (4)$^{-}$ & 7.4 \textit{(4)} & 69 & 0.40 \textit{(14)} & 0.0195 \textit{(38)} \\ 
 65.13 & (3)$^{-}$ & 4.8 \textit{(3)} & 69 & 0.14 \textit{(10)} & 0.0046 \textit{(29)} \\ 
 76.15 & 4$^{-}$ & 19.7 \textit{(6)} & 74 \textit{(5)} & 0.208 \textit{(34)} & 
0.0224 \textit{(34)} \\ 
 79.89 & 4$^{-}$ & 4.2 \textit{(3)} & 69 & 0.27 \textit{(14)} & 0.0079 \textit{(
30)} \\ 
 83.775 & 3$^{-}$ & 65.8 \textit{(15)} & 76 \textit{(5)} & 3.55 \textit{(16)} & 
0.772 \textit{(16)} \\ 
 94.90 & (4$^{-}$) & 5.6 \textit{(4)} & 69 & $<$ 0.018 & 
 $<$ 0.00065 \\ 
 99.54 & 4$^{-}$ & 263 \textit{(4)} & 79 \textit{(5)} & 0.034 \textit{(12)} & 
0.0142 \textit{(51)} \\ 
 102.80 & 3$^{-}$ & 173.6 \textit{(31)} & 76 \textit{(7)} & 1.486 \textit{(70)}
& 0.470 \textit{(15)} \\ 
 107.06 & 4$^{-}$ & 49.7 \textit{(16)} & 82 \textit{(5)} & 1.100 \textit{(81)} & 
0.217 \textit{(11)} \\ 
 108.58 & 4$^{-}$ & 1.0 \textit{(4)} & 69 & $<$ 1.2 & $<$ 0.0077 \\ 
 123.95 & 3$^{-}$ & 151.5 \textit{(33)} & 73 \textit{(6)} & 1.276 \textit{(71)}
& 0.392 \textit{(17)} \\ 
 140.30 & 3$^{-}$ & 77.7 \textit{(21)} & 69 & 0.78 \textit{(13)} & 0.192 \textit{(14)} \\ 
 143.27 & 4$^{-}$ & 3.6 \textit{(5)} & 69 & $<$ 0.72 & 
$<$ 0.017 \\ 
 151.54 & 3$^{-}$ & 144 \textit{(4)} & 75 \textit{(5)} & 0.528 \textit{(52)} & 
0.159 \textit{(14)} \\ 
 161.03 & 3$^{-}$ & 47.6 \textit{(21)} & 69 & 4.04 \textit{(82)} & 0.780 \textit{(48)} \\ 
 161.88 & 4$^{-}$ & 15.6 \textit{(12)} & 69 & 5.6 \textit{(17)} & 0.523 \textit{(48)} \\ 
 163.62 & 4$^{-}$ & 175 \textit{(4)} & 77 \textit{(4)} & 0.342 \textit{(59)} & 
0.129 \textit{(22)} \\ 
 171.80 & 4$^{-}$ & 18.5 \textit{(11)} & 69 \textit{(4)} & 0.35 \textit{(11)} & 
0.038 \textit{(11)} \\ 
 179.68 & 3$^{-}$ & 9.0 \textit{(9)} & 69 & 1.92 \textit{(67)} & 0.109 \textit{(
18)} \\ 
 184.76 & 3$^{-}$ & 356 \textit{(6)} & 69 & 20.9 \textit{(11)} & 7.83 \textit{(
10)} \\ 
 191.07 & 3$^{-}$ & 31.5 \textit{(16)} & 79 \textit{(5)} & 3.09 \textit{(30)} & 
0.423 \textit{(30)} \\ 
 193.61 & 4$^{-}$ & 5.6 \textit{(10)} & 69 & $<$ 0.53 & 
$<$ 0.019 \\ 
 198.03 & 3$^{-}$ & 13.7 \textit{(12)} & 61 \textit{(4)} & $<$ 0.034
& $<$ 0.0030 \\ 
 206.03 & 4$^{-}$ & 207 \textit{(5)} & 83 \textit{(5)} & $<$ 0.043 & 
$<$ 0.017 \\ 
 221.03 & 3$^{-}$ & 118 \textit{(4)} & 67 \textit{(6)} & 0.98 \textit{(16)} & 
0.286 \textit{(46)} \\ 
 222.68 & 3$^{-}$ & 224 \textit{(6)} & 86 \textit{(5)} & 2.47 \textit{(20)} & 
0.808 \textit{(58)} \\ 
 225.91 & (3$^{-}$) & 2.9 \textit{(13)} & 69 & 11.1 \textit{(63)} & 0.222 \textit{(31)} \\ 
 228.48 & 4$^{-}$ & 1.7 \textit{(4)} & 69 & $<$ 1.8 & $<$ 0.021 \\ 
 240.76 & 4$^{-}$ & 19.1 \textit{(18)} & 91 \textit{(6)} & 0.37 \textit{(18)} & 
0.033 \textit{(16)} \\ 
 247.62 & 4$^{-}$ & 163 \textit{(6)} & 69 \textit{(6)} & 0.137 \textit{(52)} & 0.052 \textit{(19)} \\ 
 257.13 & 3$^{-}$ & 82 \textit{(4)} & 69 & 0.60 \textit{(11)} & 0.152 \textit{(26)} \\ 
 263.57 & 3$^{-}$ & 65 \textit{(4)} & 69 & 0.41 \textit{(13)} & 0.092 \textit{(
26)} \\ 
 266.26 & 4$^{-}$ & 204 \textit{(7)} & 72 \textit{(6)} & $<$ 0.17 & 
$<$ 0.069 \\ 
 270.72 & 3$^{-}$ & 76 \textit{(4)} & 85 \textit{(6)} & 0.97 \textit{(16)} & 0.213 \textit{(34)} \\ 
 274.40 & 3$^{-}$ & 19.1 \textit{(23)} & 69 & 2.65 \textit{(82)} & 0.279 \textit{(36)} \\ 
 283.28 & 4$^{-}$ & 22.8 \textit{(25)} & 58 \textit{(10)} & $<$ 0.14 & 
$<$ 0.021 \\ 
 290.10 & (4)$^{-}$ & 41.3 \textit{(33)} & 68 \textit{(6)} & 0.53 \textit{(15)} & 0.105 \textit{(27)} \\ 
 308.30 & 3$^{-}$ & 8.3 \textit{(20)} & 69 & $<$ 0.51 & 
$<$ 0.026 \\ 
 312.06 & 4$^{-}$ & 27.6 \textit{(26)} & 69 & 0.41 \textit{(22)} & 0.060 \textit{(29)} \\ 
 321.13 & 3$^{-}$ & 11.4 \textit{(11)} & 69 & $<$ 0.46 & 
$<$ 0.031 \\ 
330.10 & 3$^{-}$ & 67 \textit{(4)} & 69 & 0.41 \textit{(22)} & 0.095 \textit{(47)} \\ 
332.10 & 4$^{-}$ & 73 \textit{(4)} & 69 & $<$ 0.39 & $<$ 0.10 \\ 
340.4 & 4$^{-}$ & 178 \textit{(7)} & 69 & 0.18 \textit{(11)} & 0.070 \textit{(41)} \\ 
349.86 & 3$^{-}$ & 68 \textit{(4)} & 69 & 0.38 \textit{(19)} & 0.088 \textit{(43)} \\ 
359.32 & 4$^{-}$ & 402 \textit{(12)} & 69 & 0.28 \textit{(11)} & 0.131 \textit{(52)} \\ 
362.15 & 4$^{-}$ & 31 \textit{(4)} & 69 & $<$ 0.49 & $<$ 0.077 \\ 
379.2 & 4$^{-}$ & 393 \textit{(12)} & 69 & 0.72 \textit{(20)} & 0.340 \textit{(90)} \\ 
382.4 & 3$^{-}$ & 139 \textit{(8)} & 69 & 1.28 \textit{(44)} & 0.39 \textit{(13)} \\ 
385.16 & 4$^{-}$ & 122 \textit{(7)} & 69 & 3.70 \textit{(67)} & 1.27 \textit{(13)} \\ 
396.5 & (4)$^{-}$ & 67 \textit{(5)} & 69 & 0.54 \textit{(33)} & 0.141 \textit{(81)} \\ 
398.6 & 3$^{-}$ & 109 \textit{(7)} & 69 & 0.41 \textit{(24)} & 0.116 \textit{(67)} \\ 
405.1 & 3$^{-}$ & 34 \textit{(4)} & 69 & 1.13 \textit{(48)} & 0.178 \textit{(61)} \\ 
412.0 & 3$^{-}$ & 55 \textit{(5)} & 69 & $<$ 0.49 & $<$ %
0.10 \\ 
418.3 & (4)$^{-}$ & 235 \textit{(12)} & 69 & 0.43 \textit{(18)} & 0.180 \textit{(74)} \\ 
421.8 & 4$^{-}$ & 68 \textit{(5)} & 69 & 0.73 \textit{(32)} & 0.193 \textit{(75)} \\ 
433.1 & (3$^{-}$) & 17 \textit{(4)} & 69 & 2.8 \textit{(17)} & 0.26 \textit{(13)}
\\ 
435.7 & 3$^{-}$ & 154 \textit{(9)} & 69 & 1.94 \textit{(63)} & 0.61 \textit{(18)} \\ 
439.5 & 4$^{-}$ & 40 \textit{(5)} & 69 & 9.5 \textit{(26)} & 1.81 \textit{(17)}
\\ 
446.9 & 3$^{-}$ & 7 \textit{(3)} & 69 & $<$ 1.59 & $<$ 0.067 \\ 
458.6 & 4$^{-}$ & 100 \textit{(7)} & 69 & 0.24 \textit{(16)} & 0.077 \textit{(50)} \\ 
462.9 & 3$^{-}$ & 53 \textit{(6)} & 69 & $<$ 0.30 & $<$ %
0.061 \\ 
476.0 & 4$^{-}$ & 117 \textit{(8)} & 69 & 0.41 \textit{(25)} & 0.140 \textit{(82)} \\ 
479.8 & 3$^{-}$ & 177 \textit{(11)} & 69 & 1.43 \textit{(39)} & 0.47 \textit{(11)} \\ 
486.4 & 3$^{-}$ & 111 \textit{(8)} & 69 & 1.70 \textit{(43)} & 0.48 \textit{(10)} \\ 
496.2 & 4$^{-}$ & 120 \textit{(9)} & 69 & 0.28 \textit{(20)} & 0.096 \textit{(68)} \\ 
498.6 & (3)$^{-}$ & 294 \textit{(15)} & 69 & 0.41 \textit{(23)} & 0.150 \textit{(83)} \\ 
518.2 & 4$^{-}$ & 474 \textit{(20)} & 69 & 1.92 \textit{(30)} & 0.93 \textit{(13)} \\ 
528.9 & 4$^{-}$ & 72 \textit{(7)} & 69 & 0.34 \textit{(23)} & 0.093 \textit{(60)} \\ 
532.5 & 3$^{-}$ & 60 \textit{(7)} & 69 & 0.42 \textit{(30)} & 0.091 \textit{(62)} \\ 
538.1 & 4$^{-}$ & 575 \textit{(22)} & 69 & 0.64 \textit{(23)} & 0.32 \textit{(11)} \\ 
546.0 & (3)$^{-}$ & 185 \textit{(12)} & 69 & 0.85 \textit{(34)} & 0.28 \textit{(11)} \\ 
553.2 & 3$^{-}$ & 367 \textit{(26)} & 69 & 0.33 \textit{(22)} & 0.125 \textit{(83)} \\ 
554.5 & 4$^{-}$ & 248 \textit{(20)} & 69 & 1.77 \textit{(45)} & 0.76 \textit{(17)} \\ 
559.7 & 3$^{-}$ & 207 \textit{(14)} & 69 & 0.82 \textit{(44)} & 0.28 \textit{(15)} \\ 
563.4 & 4$^{-}$ & 219 \textit{(15)} & 69 & 1.39 \textit{(55)} & 0.58 \textit{(22)} \\ 
567.6 & (3$^{-}$) & 38 \textit{(7)} & 69 & 4.5 \textit{(17)} & 0.76 \textit{(19)} \\ 
574.3 & 4$^{-}$ & 101 \textit{(9)} & 69 & 2.37 \textit{(64)} & 0.75 \textit{(15)} \\ 
580.2 & 3$^{-}$ & 124 \textit{(11)} & 69 & 0.29 \textit{(20)} & 0.085 \textit{(57)} \\ 
587.8 & 3$^{-}$ & 83 \textit{(9)} & 69 & 0.95 \textit{(46)} & 0.24 \textit{(11)}
\\ 
597.4 & 4$^{-}$ & 176 \textit{(13)} & 69 & 0.52 \textit{(27)} & 0.20 \textit{(10)} \\ 
606.0 & 4$^{-}$ & 126 \textit{(11)} & 69 & 0.56 \textit{(32)} & 0.19 \textit{(11)} \\ 
612.6 & (3$^{-}$) & 93 \textit{(10)} & 69 & 1.13 \textit{(59)} & 0.30 \textit{(15)} \\ 
617.2 & (3)$^{-}$ & 493 \textit{(25)} & 69 & $<$ 0.97 & 
$<$ 0.38 \\ 
622.6 & (3$^{-}$) & 151 \textit{(13)} & 69 & 0.66 \textit{(41)} & 0.21 \textit{(13)} \\ 
625.3 & (4$^{-}$) & 74 \textit{(10)} & 69 & $<$ 1.1 & $<$ 0.28 \\ 
634.0 & 3$^{-}$ & 29 \textit{(8)} & 69 & 1.8 \textit{(1.1)} & 0.26 \textit{(12)} \\
644.7 & (3$^{-}$) & 60 \textit{(9)} & 69 & 0.92 \textit{(56)} & 0.20 \textit{(11)} \\ 
648.5 & (3$^{-}$) & 209 \textit{(15)} & 69 & $<$ 0.40 & 
$<$ 0.14 \\ 
651.9 & (4$^{-}$) & 102 \textit{(11)} & 69 & 2.16 \textit{(77)} & 0.69 \textit{(21)} \\ 
659.4 & (3)$^{-}$ & 80 \textit{(10)} & 69 & 5.9 \textit{(16)} & 1.48 \textit{(26)} \\
667.0 & 4$^{-}$ & 65 \textit{(10)} & 69 & 15.6 \textit{(41)} & 4.00 \textit{(34)} \\ 
677.5 & (3$^{-}$) & 159 \textit{(14)} & 69 & $<$ 0.48 & 
$<$ 0.15 \\ 
683.1 & (4$^{-}$) & 236 \textit{(18)} & 69 & $<$ 0.29 & 
$<$ 0.12 \\ 
687.4 & (3$^{-}$) & 19 \textit{(9)} & 69 & $<$ 2.6 & $<$ 0.24 \\ 
697.0 & (4)$^{-}$ & 87 \textit{(12)} & 69 & $<$ 1.5 & $<$ 0.44\\ \hline\hline 
\end{longtable}

The accuracy of the extracted $\alpha $ widths depends on the accuracy of
the $J^{\pi }$, $\Gamma _{n}$, and $\Gamma _{\gamma }$ assignments for the
resonances because our measurement technique determines only the resonance
areas, $A_{\alpha }=g_{J}\Gamma _{\alpha }\Gamma _{n}/\Gamma $, where $g_{J}$
is the statistical factor ($g_{J}=(2J+1)/[(2I+1)(2i+1)]$, where $J$, $I$,
and $i$ are the spins of the resonance, $^{147}$Sm, and the neutron,
respectively). The uncertainties in the resonance areas in Table \ref%
{Sm147ResTable} are the one-standard-deviation uncertainties determined in
fitting the data. The uncertainties in the $\alpha $ widths include
additional contributions (added in quadrature) from the uncertainties in the
neutron and radiation widths. The uncertainties in the neutron and radiation
widths were taken from Ref. \cite{Su98}. For resonances with unknown
radiation widths a ``factor of two'' ($\pm $ 23 meV) uncertainty was
assumed, which should be a conservative overestimate of the uncertainty
because the measured radiation widths \cite{Mi81,Ge93} show very little
variability.

For 23 of the 104 resonances in this region, the fitted resonance areas had
a relative uncertainty greater than 70\%. In these cases we give only upper
limits for the resonance areas in Table \ref{Sm147ResTable} equal to the
fitted values plus the one-standard-deviation uncertainties determined by
SAMMY. The upper limits on the $\alpha $ widths given in Table \ref%
{Sm147ResTable} for these cases were calculated from the upper limits on the
resonance areas using the listed spins and neutron and radiation widths. The
uncertainties in the neutron and radiation widths have negligible effects in
these cases and hence were not taken into account when calculating these $%
\alpha $-width upper limits.

\subsection{Average resonance parameters}

For comparison to statistical model calculations, distributions or averages
of resonance parameters are needed. Some of these quantities were determined
from a resonance analysis of the total cross section \cite{Mi81}. With the
resonance spin and $\gamma $-width information from Ref. \cite{Ge93} and the 
$\alpha $-width information from this work, it is possible to calculate
these quantities for each of the two possible spin states for \textit{s}%
-wave resonances and to include the $\alpha $ channels as well as the
neutron channels. Useful quantities for comparison to statistical model
calculations include strength functions, $S$, average level spacings, $D$,
average widths, $<\Gamma >$, and the distributions of widths (see Sec.\ \ref%
{StatModComp} for a detailed discussion of these quantities). Care must be
taken to account for effects due to missed resonances.

\subsubsection{Neutron and gamma channels}

\begin{figure}[tbp]
\includegraphics*[width=90mm,keepaspectratio] {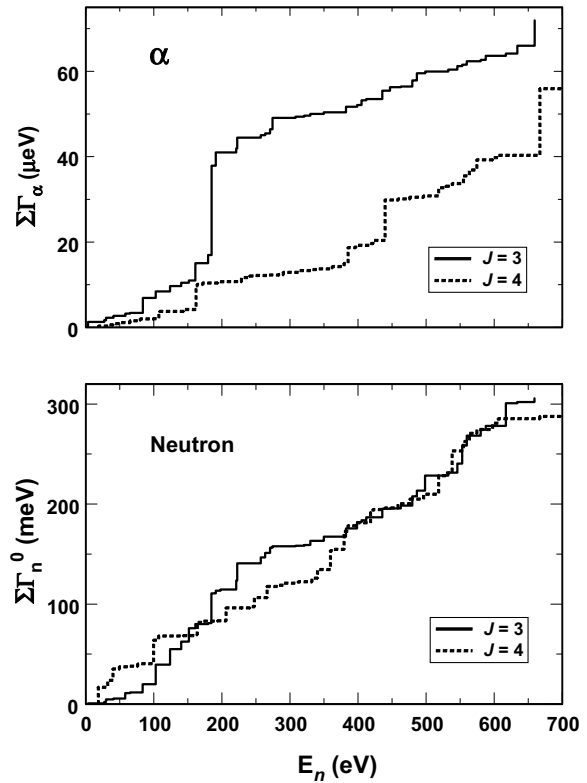}
\caption{Strength function plots for the $\protect\alpha $ (top) and neutron
(bottom) channels. Solid and dashed staircase plots represent results for 3$%
^{-}$ and 4$^{-}$ resonances, respectively. The measured strength functions
are equal to the slopes of these staircase plots. }
\label{CumuAlphAndNeutWidFig}
\end{figure}

The \textit{s}-wave neutron strength function, $S_{0}=<\Gamma
_{n}^{0}>/D_{0} $, where $<\Gamma _{n}^{0}>$ is the average \textit{s}-wave
reduced neutron width ($\Gamma _{n}=\Gamma _{n}^{0}\times \sqrt{E_{n}}$) and 
$D_{0}$ is the average \textit{s}-wave level spacing, can be determined from
the slope of a plot of the cumulative reduced neutron width versus resonance
energy resulting from a resonance analysis of total cross section
measurements. This well-known technique for determining the neutron strength
function is relatively insensitive to missing resonances because only
resonances having small neutron widths are expected to be missed in total
cross section measurements and the slope of the cumulative reduced neutron
width versus resonance energy is little affected by missing resonances with
small widths. In Ref. \cite{Mi81}, this technique was used to calculate $%
10^{4}S_{0}=4.8\pm 0.5$. Using the resonance spin information from Ref. \cite%
{Ge93} together with the neutron widths from Ref. \cite{Mi81}, strength
function plots for the two possible \textit{s}-wave spins were constructed
and are shown in Fig. \ref{CumuAlphAndNeutWidFig}. Linear fits to the data
in Fig. \ref{CumuAlphAndNeutWidFig} yield strength functions of $(4.6\pm
1.0)\times 10^{-4}$ and $(4.3\pm 0.9)\times 10^{-4}$ for $J=3$ and $4$
resonances, respectively, in agreement with the combined \textit{s}-wave
strength function of Ref. \cite{Mi81}. Uncertainties in the strength
functions were estimated from the number of observed resonances as described
in Ref. \cite{Mu81}.

In Ref. \cite{Mi81}, the well known technique of determining the level
spacing from the inverse of the slope of the cumulative number of resonances
versus resonance energy (in the lower energy region where missed resonances
are insignificant) was used to calculate an average level spacing $%
D_{0}=5.7\pm 0.5$ eV for all \textit{s}-wave resonances. Using this level
spacing and assuming that the number of resonances for each spin is
proportional to $2J+1$, leads to $N=54$ and $69$ resonances by 700 eV for $%
J=3$ and $4$ resonances, respectively (implying $D_{0}=13.0$ and $10.1$ eV
for $J=3$ and $4$ resonances, respectively). These level spacings, together
with the above strength functions for $J=3$ and $4$, yield $<\Gamma
_{n}^{0}>=5.98$ and $4.4$ meV, for $J=3$ and $4$ resonances, respectively.

\begin{figure}[tbp]
\includegraphics*[width=90mm,keepaspectratio] {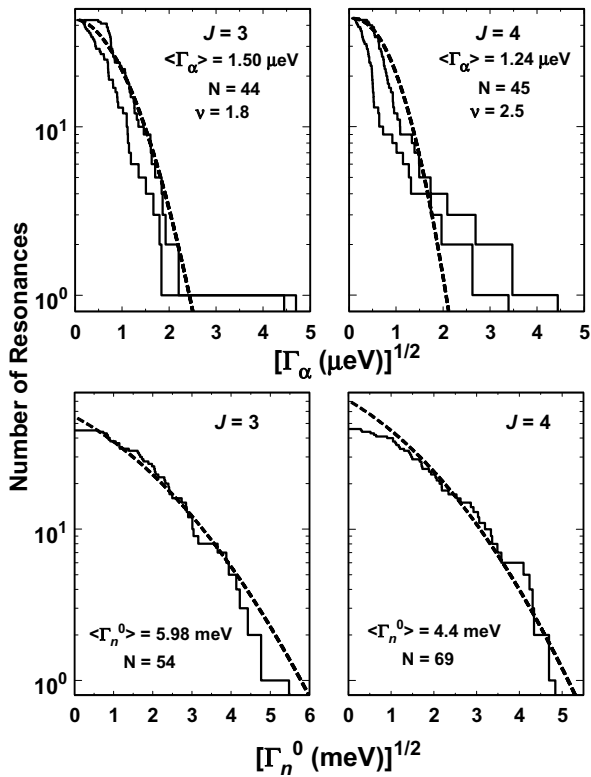}
\caption{Integral distributions of total $\protect\alpha $ (top) and reduced
neutron (bottom) widths for 3$^{-}$ (left) and 4$^{-}$ (right) resonances in 
$^{147}$Sm. The histograms show the measured number of resonances having
widths greater than a given width versus that width. In an attempt to depict
the experimental uncertainties, histograms for the $\protect\alpha $ widths
are plotted at the measured widths plus and minus their respective
uncertainties. The dashed curves show the expected $\protect\chi ^{2}$
distributions for the average widths, number of resonances, and degrees of
freedom ($\protect\nu =1$ for neutrons) indicated.}
\label{PTDistAlAndNeutFig}
\end{figure}

Neutron widths are expected to obey Porter-Thomas \cite{Po56} distributions
having average widths and numbers of resonances consistent with the above
strength function and level spacing determinations. This was found to be the
case in Ref. \cite{Mi81} for all (sum of $J=3$ and $4$) \textit{s}-wave
resonances. The separate neutron width distributions for $J=3$ and $4$
resonances are shown in Fig. \ref{PTDistAlAndNeutFig}. We will discuss how
well they compare to the expected Porter-Thomas distributions in Section \ref%
{PossNonStat}.

\subsubsection{$\protect\alpha$ channel}

\begin{figure}[tbp]
\includegraphics*[width=90mm,keepaspectratio] {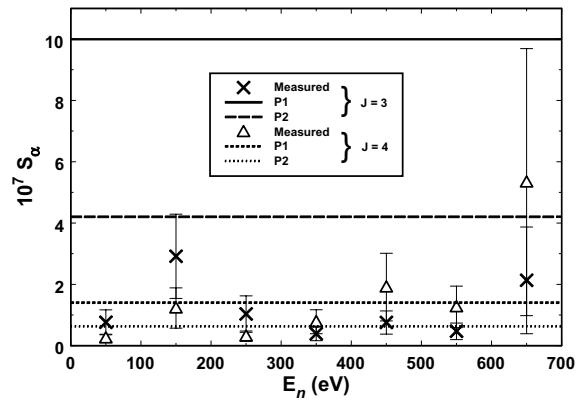}
\caption{Measured $\protect\alpha $ strength functions averaged over 100-eV
intervals for $J^{\protect\pi }=3^{-}$ (X's) and $4^{-}$ (triangles)
resonances. Strength functions for $J^{\protect\pi }=3^{-}$ resonances
calculated with the standard (P1) and modified (P2) NON-SMOKER parameters
are shown as solid and long-dashed lines, respectively, whereas
corresponding calculations for $J^{\protect\pi }=4^{-}$ resonances are shown
as short-dashed and dotted lines, respectively.}
\label{AvgAlStrFun}
\end{figure}

Our measurement technique is sensitive to resonance areas, $A_{\alpha
}=g_{J}\Gamma _{\alpha }\Gamma _{n}/\Gamma $; hence, missed resonances can
have small as well as large (e.g. when $\Gamma _{n}$ is small) $\alpha $
widths. Therefore, determining the $\alpha $ strength functions from the
slopes of the cumulative $\alpha $ widths versus neutron energy (top part of
Fig. \ref{CumuAlphAndNeutWidFig}), will likely result in values that are
systematically small. On the other hand, because missed resonances should
have a range of $\alpha $ widths, the average $\alpha $ widths should be
less affected by associated systematic effects and strength functions
calculated from the average widths should be more reliable. There is the
additional complication that the data indicate that the average $\alpha $
widths as well as the $\alpha $ strength functions show considerable
variations as functions of energy. For example, as shown in Fig. \ref%
{CumuAlphAndNeutWidFig}, there are large steps in the slopes of the
cumulative $\alpha $ widths versus neutron energy. For this reason, the $%
\alpha $ strength functions were calculated over different energy intervals.
As shown in Fig. \ref{AvgAlStrFun}, the $\alpha $ strength functions
calculated from the $\alpha $ widths ($S=<\Gamma >/D$) averaged over 100-eV
intervals show considerable variation. In addition, as shown in Fig. \ref%
{AvgAlStrFunRat}, the ratio of the $\alpha $ strength functions for the two
different \textit{s}-wave spins states changes dramatically near 300 eV.
Because it is possible that this effect is caused by incorrect spin
assignments and because the strength functions as well as their ratio are
relatively constant in the 0-300 eV and 300-600 eV intervals, we list the
strength functions calculated for these two ranges as well as for the entire
0-600 eV interval in Table \ref{AlphaStrengthTable}. We did not include the
600-700 keV interval because there are only about half as many resonances
having firm $J^{\pi }$ assignments in this energy range as in the six lower
energy bins. We compare these values to statistical model calculations in
Section \ref{StatModComp}.

Distributions of $\alpha $ widths for the two \textit{s}-wave spin states
are shown in Fig. \ref{PTDistAlAndNeutFig}. Alpha widths for transitions to
individual final states are expected to obey Porter-Thomas distributions,
but we were unable to resolve individual $\alpha $ groups because our
samples were too thick. Therefore, the $\alpha $ width distributions are
convolutions of Porter-Thomas distributions. This will be discussed more
fully in Section \ref{PossNonStat}.

\begin{figure}[tbp]
\includegraphics*[width=90mm,keepaspectratio] {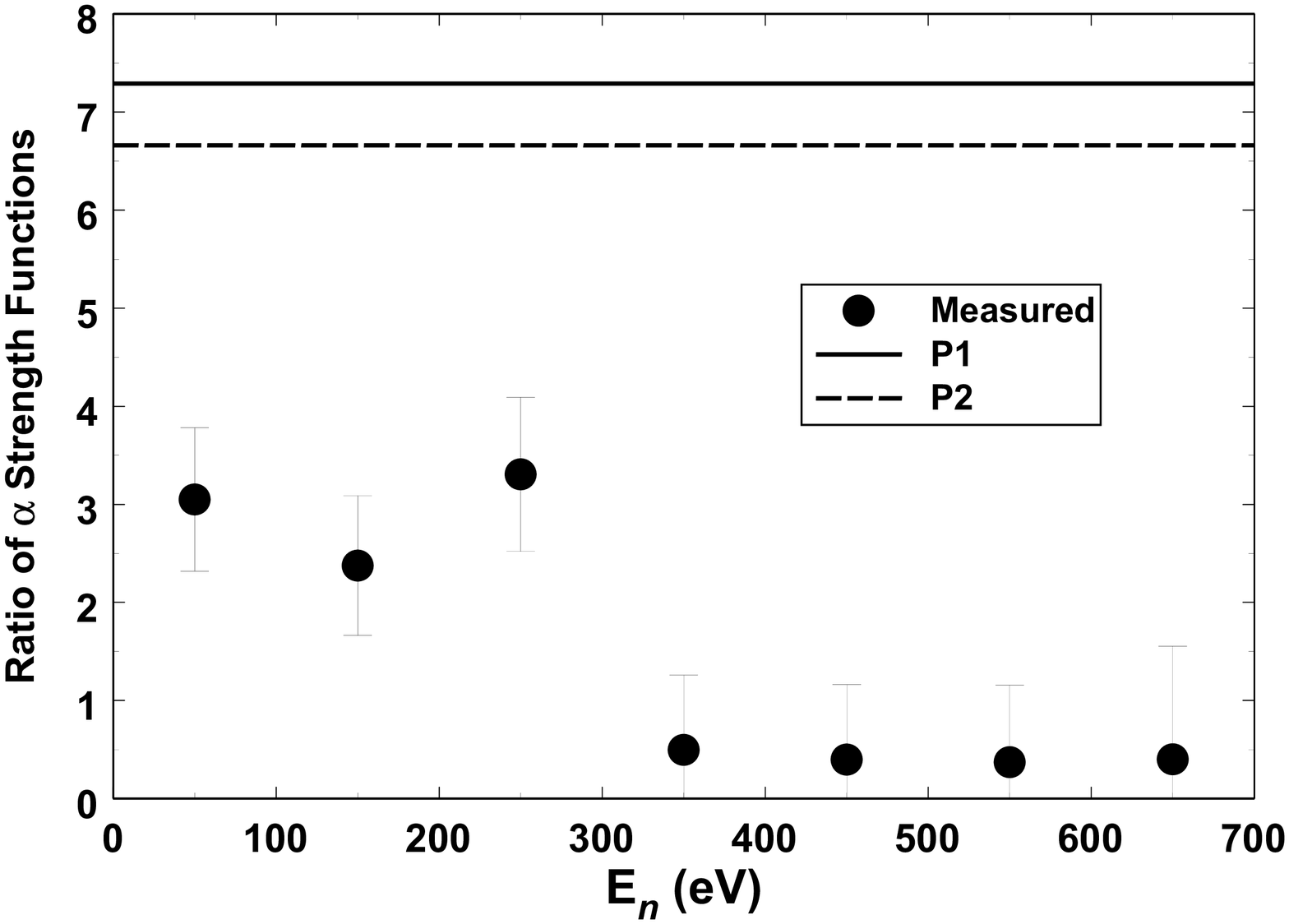}
\caption{Ratios ($J^{\protect\pi }=3^{-}$ to $4^{-}$ resonances) of $\protect%
\alpha $ strength functions averaged over 100-eV intervals. Measured ratios
are shown as solid circles. Ratios calculated with the standard (P1) and
modified (P2) NON-SMOKER parameters are shown as solid and long-dashed
lines, respectively.}
\label{AvgAlStrFunRat}
\end{figure}

\begin{table*}[tbp]
\caption{Neutron and $\protect\alpha $ strength functions from experiment
and optical models.}
\label{AlphaStrengthTable}\centering%
\begin{ruledtabular}
\begin{tabular}{cccccccc}
    & \multicolumn{2}{c}{Neutrons} & \multicolumn{5}{c}{Alphas} \\
$J$ & \multicolumn{2}{c}{10$^{4}S_{0}$} & \multicolumn{5}{c}{10$%
^{6}S_{\alpha }$} \\ \hline
& Experiment & Optical Model & \multicolumn{3}{c}{Experiment} & \multicolumn{2}%
{c}{Optical Model} \\ 
&  &  & 0-600 eV & 0-300 eV & 300-600 eV & P1 & P2 \\  \hline
3 & $4.6\pm 1.0$ & 9.2 & $0.116\pm 0.025$ & $0.171\pm 0.052$ & $0.0560\pm 0.018$ & 1.0 & 0.42 \\ 
4 & $4.3\pm 0.9$ & 9.2 & $0.092\pm 0.020$ & $0.058\pm 0.018$ & $0.127\pm 0.039$ & 0.14 & 0.063\\ 
\end{tabular}%
\end{ruledtabular}
\end{table*}

\section{Comparison to previous work}

There is reasonably good agreement between the $\alpha $ widths we
determined and previous work (as compiled in Ref. \cite{Su98}) except that
we have many more measured widths and hence many fewer upper limits. There
are however at least two important differences between previous results and
the resonance parameters reported herein. First, the neutron and gamma
widths used to extract the $\alpha $ widths typically were not reported in
previous work. Because the $\alpha $ widths extracted can vary substantially
depending on the values of these other parameters, comparisons to previously
reported $\alpha $ widths without this information are of limited value.
Second, originally \cite{Gl2002}, we followed the example of Ref. \cite{An84}
and assigned the stronger peaks in our data to the same resonances they used
even though, as the energy increased, our energy scale indicated that the
peaks actually corresponded to the next higher-energy resonances. As a
result, our resonance energies became increasing larger than those of Ref. %
\cite{Su98} as the energy increased. For example, with this scheme the
resonance we observed at 439.5 eV would correspond to the resonance in Ref. %
\cite{Su98} at 435.7 eV rather than to the much closer resonance at 440.2
eV. After a recheck of our energy calibration, using the present data as
well as data from $^{143}$Nd(\textit{n},$\alpha $) \cite{Ko2001} and $^{95}$%
Mo(\textit{n},$\alpha $) \cite{Ra2003} measurements run under the same
conditions using the same apparatus,\ we decided that it was not possible
for our energy scale to be wrong in this manner. Therefore, we assigned the
resonances we observed to the closest ones in Ref. \cite{Su98}. As a result,
many of the spin assignments as well as the extracted $\alpha $ widths
(because the widths were calculated using the wrong spins and neutron and
radiation widths) in our preliminary report \cite{Gl2002} are incorrect, and
several resonances with large $\alpha $ widths at the higher energies are
now assigned to $J^{\pi }$= 4$^{-}$ rather than $J^{\pi }$= 3$^{-}$.

\begin{figure}[tbp]
\includegraphics*[width=90mm,keepaspectratio] {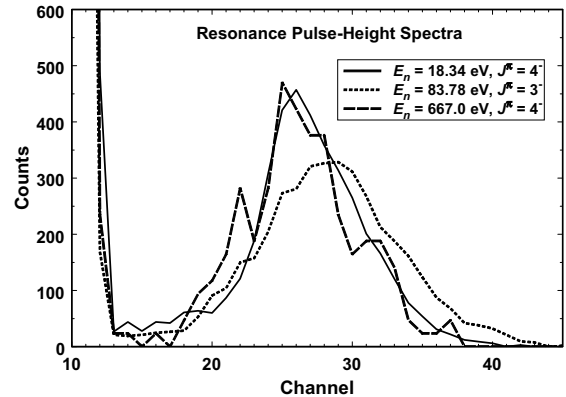}
\caption{Pulse-height spectra from the $^{147}$Sm(\textit{n},$\protect\alpha 
$) reaction for the 18.340-, 83.775-, and 667.0-eV resonances. The spectra
have been normalized to have equal areas.}
\label{J3CompJ4Fig}
\end{figure}

Further evidence that our spin assignments are correct can be seen in the
pulse-height distributions for the resonances. As illustrated in Fig. \ref%
{EnergyLevelDiagFig}, because parity must be conserved, 4$^{-}$ resonances
in $^{148}$Sm are forbidden from decaying to the 0$^{+}$ ground state of $%
^{144}$Nd by $\alpha $ emission whereas 3$^{-}$ resonances are not.
Therefore, the pulse-height spectrum for $\alpha $ particles from 4$^{-}$
resonances will not include the highest-energy group corresponding to decay
to the ground state of $^{144}$Nd. Because our samples were relatively
thick, we could not resolve the various $\alpha $-particle groups. However,
as shown in Fig. \ref{J3CompJ4Fig}, there is a discernible difference
between the pulse-height spectra for unambiguously assigned $J^{\pi }$= 3$%
^{-}$ and 4$^{-}$ resonances. As expected, the $J^{\pi }$= 3$^{-}$ resonance
at 83.775 eV occurs at larger pulse height than the $J^{\pi }$= 4$^{-}$
resonance at 18.340 eV. This figure also demonstrates that the pulse height
spectrum corresponding to the resonance at 667.0 eV clearly favors a 4$^{-}$
assignment for this resonance. Similar comparisons show that 4$^{-}$
assignments also are favored for several other resonances with fairly large $%
\alpha $ widths which had been associated with 3$^{-}$ resonances in
previous work \cite{An84}, lending confidence that the energy calibration,
and therefore the spin assignments, of the present work are correct.

\begin{figure}[tbp]
\includegraphics*[width=90mm,keepaspectratio] {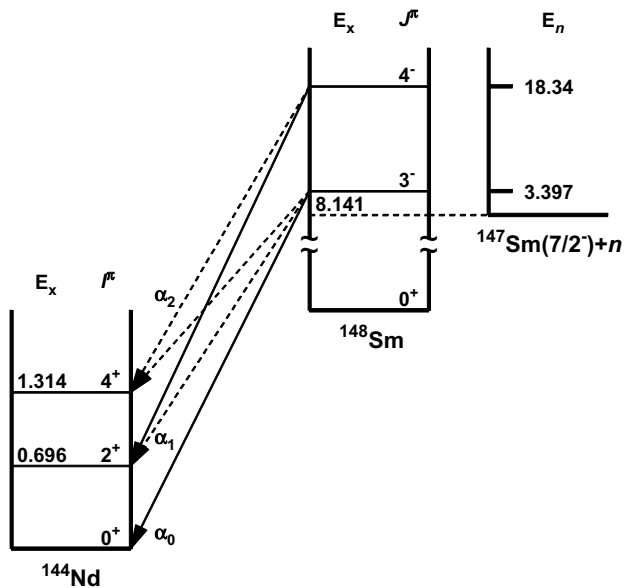}
\caption{Energy-level diagrams depicting the $^{147}$Sm(\textit{n},$\protect%
\alpha $)$^{144}$Nd reaction ($Q=10.127$ MeV). Excitation energies $E_{x}$
are given in MeV whereas laboratory neutron energies $E_{n}$ are given in
eV. The energy scales of the $^{144}$Nd and $^{148}$Sm parts of the figure
differ by a factor of 17 million. $J^{\protect\pi }$ = 3$^{-}$ levels in $%
^{148}$Sm can $\protect\alpha $ decay to the ground as well as excited (with
substantially reduced penetrability) states of $^{144}$Nd. $J^{\protect\pi }$
= 4$^{-}$ levels in $^{148}$Sm are parity-forbidden from decaying to the
ground state of $^{144}$Nd, but can decay to excited states.}
\label{EnergyLevelDiagFig}
\end{figure}

\section{Possible non-statistical effects\label{PossNonStat}}

Analysis of previous $^{147}$Sm($n,\alpha $) measurements \cite%
{Po72,Ba76,An84} have hinted at the possible presence of non-statistical
effects. The improved precision and accuracy of the current measurements
together with improved resonance parameter information \cite{Ge93} makes it
possible to perform more robust tests for non-statistical effects.

\begin{figure}[tbp]
\includegraphics*[width=90mm,keepaspectratio] {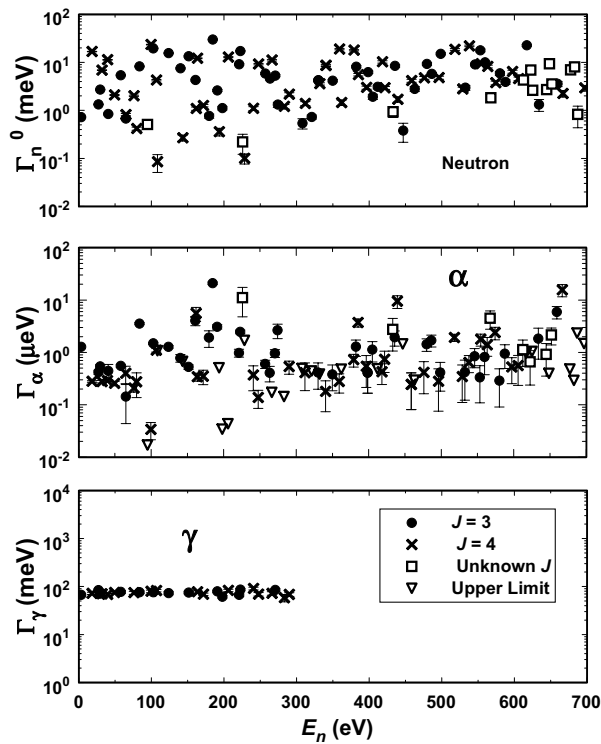}
\caption{Reduced neutron (top), $\protect\alpha $ (middle), and $\protect%
\gamma $ (bottom) widths for $^{147}$Sm resonances versus neutron energy.
The reduced neutron and $\protect\gamma $ widths are from previous work as
compiled in Ref. \protect\cite{Su98} whereas the $\protect\alpha $ widths
are from the present work. Measured widths with $J=3$, $4$, or unknown spin
assignments are shown as circles, Xs, and squares, respectively. $\protect%
\alpha$ widths for which only upper limits were determined are plotted as
triangles.}
\label{Sm147GaGn0AndGgvsE}
\end{figure}

In nuclear statistical theory, partial widths $\Gamma $ associated with the
decay of compound nuclear states are assumed to be described by $\chi ^{2}$
distributions with $\nu $ degrees of freedom:

\begin{equation}
P(x,\nu )=\frac{\nu }{2G(\nu /2)}(\frac{\nu x}{2})^{\nu /2-1}\exp (-\frac{%
\nu x}{2})  \label{Chi2DistEq}
\end{equation}

where $G$ is the gamma function, $x=\frac{\Gamma }{<\Gamma >}$, $<\Gamma >$
is the average width, and the distribution is characterized by a dispersion
equal to $2<\Gamma >^{2}/\nu $. Neutron widths as well as partial $\alpha $
widths for specific angular momentum values to individual final states are
expected to have $\nu =1$, and hence obey Porter-Thomas \cite{Po56}
distributions. On the other hand, $\gamma $ widths show much smaller
fluctuations because the much larger number of decay channels in the $\gamma 
$-ray cascade implies a $\chi ^{2}$ distribution with many more degrees of
freedom. In our present experiment, we were unable to resolve the individual 
$\alpha $ groups, and hence only the total $\alpha $ widths summed over the
various possible final states were measured. Because the partial $\alpha $
widths fluctuate independently, the fluctuations of the total $\alpha $
widths, $\Gamma _{\alpha }=\sum \Gamma _{\alpha c}$, are expected to be
smaller owing to random cancellations of the partial widths. Therefore, the
distributions of total $\alpha $ widths are expected to be narrower ($\nu >1$%
) than for the partial widths. The distribution for total $\alpha $ widths
are complicated convolutions of partial distributions with $\nu =1$ and
different average values $<\Gamma _{\alpha c}>$. Using a Monte Carlo method,
it has been shown \cite{Po70} that this convolution does not have to be
calculated but that instead equation \ref{Chi2DistEq}, after the partial
width $\Gamma _{\alpha c}$ and its average $<\Gamma _{\alpha c}>$ are
replaced by their corresponding total values $\Gamma _{\alpha }$ and $%
<\Gamma _{\alpha }>$, can be used to describe the distribution of total $%
\alpha $ widths, albeit with an effective degrees of freedom,

\begin{equation}
\nu _{eff}=\frac{(\sum P_{c})^{2}}{\sum P_{c}^{2}}\quad .
\label{EffDegFreeEq}
\end{equation}%
where $P_{c}$ is the penetrability for $\alpha $ particles in channel $c$.
In the present case, the calculated \cite{Ba76} degrees of freedom are $\nu
= $ 1.8 and 2.5 for 3$^{-}$ and 4$^{-}$ resonances, respectively. Therefore,
it is expected that the fluctuations in the total $\alpha $ widths will be
intermediate between that for neutrons and gammas. However, as shown in Fig. %
\ref{Sm147GaGn0AndGgvsE}, although the neutron and gamma widths fluctuate as
expected, the $\alpha $ widths do not appear to follow the expected
behavior. For example, instead of being intermediate to the gamma and
neutron distributions, the $\alpha $ widths show the largest fluctuations.
Also, there appear to be regions of energy in which the $\alpha $ widths
fluctuate about as expected and other regions of much larger widths and/or
fluctuations. These effects are evident for both $J^{\pi }=3^{-}$ and $4^{-}$
resonances.

Recasting the data into plots of the cumulative widths versus resonance
energy, as shown in Fig. \ref{CumuAlphAndNeutWidFig} further illustrates
that whereas the neutron widths behave as expected, the $\alpha $ widths
show indications of non-statistical effects. The large steps in the
cumulative distributions of the $\alpha $ widths might be indications of
states in $^{148}$Sm with non-statistical properties.

Converting the data to the integral width distributions shown in Fig. \ref%
{PTDistAlAndNeutFig} allows a more quantitative comparison with theory. In
the case of neutrons, the widths are expected to follow a Porter-Thomas
distributions with average widths calculated from the measured strength
functions and level spacing.\ As shown in the bottom part of Fig. \ref%
{PTDistAlAndNeutFig}, there is good agreement between the observed and
expected width distributions for neutrons indicating that the neutron widths
behave as expected when separated according to spin. The fit to the
Porter-Thomas distribution for $J=3$ resonances can be improved by assuming
additional missed resonances and hence a smaller average reduced neutron
width for this spin. This indication that there may be slightly more $J=3$
resonances than the simple $2J+1$ assumption we used is in agreement with
the relative sizes of the level spacings predicted in NON-SMOKER (see Sec.\ %
\ref{StatModComp}) for the two \textit{s}-wave spin states.

Because missed resonances are expected to have random $\alpha $ widths,
simple averages over the measured widths should yield good estimates of the
average widths and it should be possible to compare to the expected $\chi
^{2}$ distributions without correcting for missed resonances. These
comparisons are shown in the top of Fig. \ref{PTDistAlAndNeutFig}. For the
theoretical $\chi ^{2}$ distributions, average $\alpha $ widths were
calculated from simple averages of the widths determined in the resonance
analysis and the degrees of freedom values were taken from Ref. \cite{Ba76}.
In an attempt to depict the uncertainties in the $\Gamma _{\alpha }$ values,
two histogram curves are plotted in each part of Fig. \ref%
{PTDistAlAndNeutFig} corresponding to the $\alpha $ widths plus and minus
their uncertainties. Resonances for which only upper limits were determined
are plotted at the fitted values plus and minus their uncertainties as
determined by SAMMY. For the 44 firm 3$^{-}$ resonances $<\Gamma _{\alpha }>$
=1.50 $\mu $eV, whereas for the 45 firm 4$^{-}$ resonances $<\Gamma _{\alpha
}>$ = 1.24 $\mu $eV. As shown in Fig. \ref{PTDistAlAndNeutFig}, the
theoretical distributions are substantially different from the data.
Furthermore, the agreement cannot be improved by, for example, decreasing
the average widths and increasing the number of resonances in an attempt to
correct for possible effects due to missed resonances. On the other hand,
for $J^{\pi }=3^{-}$ resonances it is possible to obtain fairly good
agreement between the data and the theoretical distribution if the resonance
with the largest $\alpha $ width at 184.76 eV is excluded (resulting in $%
<\Gamma _{\alpha }>=1.05$ eV and $N=43$). For $J^{\pi }=4^{-}$ resonances
however, reasonably good agreement can be obtained only if 3-4 resonances
with the largest $\alpha $ widths are excluded. These are the same
resonances that cause the large steps in the top part of Fig. \ref%
{CumuAlphAndNeutWidFig}.

One final piece of evidence on the unusual nature of the extracted $\alpha $
widths is revealed in a comparison of the average widths for resonances for
the two different spins. In the energy range of this work, emitted $\alpha $
particles from the $^{147}$Sm($n,\alpha $) reaction are below the Coulomb
barrier, so penetrability is a steep function of energy. Therefore, the
average $\alpha $ width for 3$^{-}$ resonances is expected to be 5-10 times
larger than for 4$^{-}$ resonances because the latter are forbidden by
parity conservation from decaying to the 0$^{+}$ ground state of $^{144}$Nd
(and hence on average have less energy) whereas the former are not. However,
as noted above, the $\alpha $ widths averaged over all resonances with firm
spin assignments are almost equal for the two spin states. Furthermore, the $%
\alpha $ strength functions for the two spins are expected to remain
constant over the range of our resonance analysis given the smallness of our
range of neutron energies compared to the energies of the emitted $\alpha $
particles ($Q_{(n,\alpha )}=10.127$ MeV). However, as can be seen in Fig. %
\ref{AvgAlStrFunRat}, our data reveal a striking disagreement with the
expectations. We find that the ratio of $\alpha $ strength functions for the
two spin states changes rather dramatically from $\approx 3$ to $\approx 0.5$
near $E_{n}=$ 300 eV.

\section{Comparison to theory\label{StatModComp}}

Even though the data show signs of non-statistical effects, it is
interesting to compare the measured average resonance parameters to those
computed from optical potentials. Although this is independent of the
Hauser-Feshbach approach, using potentials similar to those from established
statistical models allows a direct test of the potentials and an examination
of the various components contributing to the calculated cross section, and
also should yield a better understanding of non-statistical signatures.

\subsection{Definitions}

The statistical model (Hauser-Feshbach approach) \cite{hauser,gad92} assumes
the compound nucleus reaction mechanism and a nuclear level density
sufficiently high to be able to average across it. In this approach, the
cross section between a target nucleus \textit{i} and projectile \textit{j}
proceeding to exit channel \textit{e} (i.e. the $\alpha $ channel in our
case) is given by: 
\begin{widetext} 
\begin{equation}
\sigma ^{\mathrm{HF}} =\frac{\pi ^{2}}{k_{j}^{2}}\frac{(1+\delta _{ij})}{%
(2I_{i}+1)(I_{j}+1)}
\sum_{J,\pi }(2J+1)\frac{T_{j}(E,J,\pi )T_{e}(E,J,\pi )}{T^{\mathrm{%
tot}}(E,J,\pi )}W_{j,e}(E,J,\pi ).
\label{eq:hfcs}
\end{equation}%
\end{widetext}In this equation, $k_{j}$ is the wave number of the
projectile, $I_{i}$ and $I_{j}$ are the spins of the target and projectile,
respectively, and the transmission coefficients (TCs) $T(E,J,\pi )$ and
width fluctuation corrections (WFC) $W_{j,e}(E,J,\pi )$ are defined by \cite%
{gad92}: 
\begin{equation}
T(E,J,\pi )=\frac{2\pi }{D(E,J,\pi )}\left\langle \Gamma _{J,\pi
}(E)\right\rangle \quad ,
\end{equation}%
and 
\begin{eqnarray}
W_{j,e}(E,J,\pi ) &=&\left\langle \frac{\Gamma _{j,J,\pi }(E)\Gamma
_{e,J,\pi }(E)}{\Gamma _{J,\pi }^{\mathrm{tot}}(E)}\right\rangle   \notag \\
&&\times \frac{\left\langle \Gamma _{J,\pi }^{\mathrm{tot}}(E)\right\rangle 
}{\left\langle \Gamma _{j,J,\pi }(E)\right\rangle \left\langle \Gamma
_{e,J,\pi }(E)\right\rangle }.
\end{eqnarray}%
The decaying compound nucleus is characterized by the average level spacing $%
D(E,J,\pi )$ of states with spin $J$ and parity $\pi $ at the excitation
energy $E$. Note that in the laboratory $T_{j}$ contains only the transition
to the ground state of the target whereas $T_{e}$ contains a sum over all
possible transitions in the exit channel. The WFC correlate the incoming and
outgoing channels and account for pre-equilibrium effects ($W_{j,e}\leq 1$)
by rearranging the flux into different channels. In practice, they usually
are implemented by applying corrections to the calculated widths. For our
case it is important to note the difference between widths obtained with or
without WFC. The WFC are a model effect. Therefore, widths calculated 
\textit{without} WFC should be used when comparing to directly measured
widths. These are the averaged widths $\left\langle \Gamma \right\rangle $
found in the above equations. However, in the calculation of statistical
model cross sections, the WFC are important and must be included. For the (%
\textit{n},$\alpha $) energy range studied here, this strongly affects the
neutron channel, introducing a further uncertainty in the theoretical
modelling of the cross section. Nevertheless, the WFC are generally thought
to be well understood.

The strength functions $S$ extracted from the experimental data are closely
related to the theoretical TCs; 
\begin{equation}
S(E,J,\pi )=\frac{T(E,J,\pi )}{2\pi }\quad ,
\end{equation}%
and generally include all energetically and spin-algebraically allowed
transitions from the compound state $(E,J,\pi )$ to the final states. As
seen in Fig.\ \ref{EnergyLevelDiagFig}, this includes the $\alpha
_{0},\alpha _{1},\alpha _{2},\dots $ transitions in the $\alpha $ channel
for the 3$^{-}$ compound resonances but excludes the $\alpha _{0}$
transition for the 4$^{-}$ resonances. In the calculation, above the first
10 excited states in $^{144}$Nd the sum over individual states is replaced
by an integration over a level density in $^{144}$Nd. However, in the
present case most of the contribution to the total TC (or strength function)
arises from the lowest lying states included explicitly. Thus, any possible
error in the level density in the final nucleus is strongly suppressed and
does not affect the TCs.

The TC for a given transition with specified quantum numbers is computed by
solving the stationary Schr\"odinger equation with a given optical
potential. Thus, the TCs are sensitive only to the optical potential.

It becomes evident from the considerations above that, when possible, it is
preferable to compare calculated TCs to strength functions extracted from
experimental data. The TCs are primary quantities depending only on the
optical potential; thus, this approach makes it possible to disentangle the
various contributions to the theoretical uncertainty. Converting strength
functions to theoretical widths adds the uncertainty in the compound level
density $\rho (E,J,\pi )=1/D(E,J,\pi )$. Comparing cross sections includes
all possible errors in the widths of the different channels (optical
potentials), the level density, and the WFC.

\subsection{Comparison of strength functions}

Strength functions calculated employing the same methods used in the
statistical model code NON-SMOKER \cite{rau98,rau00,rau01} are compared to
the data in Figs. \ref{AvgAlStrFun} and \ref{AvgAlStrFunRat} and Table \ref%
{AlphaStrengthTable}. These calculations were made to illuminate the various
confounding effects that enter into calculations of cross sections and
reaction rates as well as to ascertain whether it is possible to reproduce
the data using optical model strength functions, especially given the
indications of non-statistical effects noted above.

Several differently parametrized potentials are available in literature, but
for simplicity we limit our investigation to one basic shape (Saxon-Woods)
and two basic parameter sets; the one of Ref. \cite{McF} (potential P1,
standard NON-SMOKER settings) and the other from Refs. \cite%
{Fro02,raufro03,Ra2003a} (potential P2). Cross sections for the $^{147}$Sm(%
\textit{n},$\alpha $)$^{144}$Nd reaction calculated using potential\ P1 are
a factor of 3.3 larger than the data \cite{Gl2000}, whereas calculations
made using potential P2 are in significantly better agreement (a factor of
1.4 higher than the data) with these data as well as data from a number of
other reactions.

Both potentials use standard Saxon-Woods shapes in the real and imaginary
parts of the radial potential $U$: 
\begin{equation}
U(r)=-\frac{V}{1+\exp \left( \frac{r-r_{\mathrm{r}}A^{1/3}}{a_{\mathrm{r}}}%
\right) }-i\frac{W}{1+\exp \left( \frac{r-r_{\mathrm{i}}A^{1/3}}{a_{\mathrm{i%
}}}\right) }\quad .
\end{equation}%
The parameters are given in Table \ref{tab:potpar}. It is interesting to
note that the imaginary parts of the two potentials are the same.
Nevertheless, P2 yields a value closer to the measured cross section. At
first glance, this seems counterintuitive as it is commonly stated that the
imaginary part of the optical potential determines the TC. However, it is
more correct to state that the TC is given by the imaginary part of the wave
function which in turn depends on the relative strengths of the real and the
imaginary parts of the potential.

\begin{table}[tbp]
\caption{Parameter of the basic Saxon-Woods potentials}
\label{tab:potpar}%
\begin{ruledtabular} 
\begin{tabular}{lrrrrrr}
Potential & \multicolumn{1}{c}{$V$} & \multicolumn{1}{c}{$r_{\mathrm{r}}$} & 
\multicolumn{1}{c}{$a_{\mathrm{r}}$} & \multicolumn{1}{c}{$W$} & 
\multicolumn{1}{c}{$r_{\mathrm{i}}$} & \multicolumn{1}{c}{$a_{\mathrm{i}}$}
\\ 
& \multicolumn{1}{c}{MeV} & \multicolumn{1}{c}{fm} & \multicolumn{1}{c}{fm}
& \multicolumn{1}{c}{MeV} & \multicolumn{1}{c}{fm} & \multicolumn{1}{c}{fm}
\\ \hline
P1 \cite{McF} & 185.0 & 1.40 & 0.52 & 25.0 & 1.4 & 0.52 \\ 
P2 \cite{Fro02,raufro03,Ra2003a} & 162.3 & 1.27 & 0.48 & 25.0 & 1.4 & 0.52%
\end{tabular}
\end{ruledtabular}
\end{table}

As can be seen, the calculated strength functions in the neutron channel are
about a factor of two larger than experiment for both $3^{-}$ and $4^{-}$
resonances. It is interesting to note that the standard NON-SMOKER \cite%
{rau00} level spacings ($D_{0}=7.4$ and $6.5$ eV for $J=3$ and $4$
resonances, respectively) are about a factor of two smaller than the
measured ones; hence, the calculated average neutron widths are in fairly
good agreement with the measured values. This illustrates the importance of
making the comparison between theory and experiment as strength functions,
for if instead average widths were compared, the confounding influence of
the level spacing might lead one to conclude that there was better agreement
between theory and experiment than there actually is.

For potential P1, in the $E_{n}=0-300$ eV range, the calculated $4^{-}$ $%
\alpha $ strength function is about a factor of two larger than measured,
but the calculated $\alpha $ strength function for $J^{\pi }=3^{-}$ is in
more serious disagreement (a factor of about 6 larger) with experiment. In
the $E_{n}=300-600$ eV range, the $4^{-}$ $\alpha $ strength function
calculated with potential P1 is in good agreement with the data, but for $%
3^{-}$ resonances the $\alpha $ strength function is in even more serious
disagreement (about a factor of 17) than it was in the lower-energy region.
Potential P2 is clearly better than P1 in predicting the $\alpha $ strength
function for $3^{-}$ resonances and for $4^{-}$ resonances in the $%
E_{n}=0-300$ eV region. However, as shown in Fig. \ref{AvgAlStrFunRat}, both
potentials strongly overpredict the $3^{-}$ to $4^{-}$ $\alpha $ strength
function ratio, potential P2 yielding a ratio only slightly smaller than
that calculated with potential P1.

These comparisons as strength functions indicate that the optical $\alpha $+$%
^{144}$Nd potential requires more adjustment than would be surmised from
comparisons as cross sections. However, it should be noted that the
cross-section comparisons were made at higher energies ($E_{n}\approx 10-500$
keV) where contributions from \textit{p}-wave resonances are expected to
become more important. Furthermore, the disagreement between calculated and
actual strength functions might become smaller at higher energies as has
been found in other reactions. Nevertheless, it is informative that both the
calculated neutron and $\alpha $ strength functions are too large - a fact
that could not be surmised from comparisons as $^{147}$Sm(\textit{n},$\alpha 
$)$^{144}$Nd cross sections.

As shown in Fig. \ref{AvgAlStrFunRat}, the ratio of the calculated $\alpha $
strength functions for the two \textit{s}-wave spin states is constant over
the energy range of resonance analysis. This is in stark contrast to the
measurements which show a steep decrease in this ratio above $E_{n}=300$ eV.
Regardless of the potential used, it will never be possible to achieve such
an abrupt change in the ratio within an optical model of strength functions.
However, it is possible that the abrupt change in the $3^{-}$ to $4^{-}$ $%
\alpha $ strength function ratio is due to incorrect spin assignments above $%
E_{n}=300$ eV. Therefore, it is interesting to study if calculations can
reproduce the measured ratio below $E_{n}=300$ eV where the data should be
very reliable. Two conclusions can be drawn from a more systematic variation
of the parameters of a Saxon-Woods $\alpha $ potential which we attempted.
Firstly, the absolute values of the strength functions are far more
sensitive to the potential than is the $3^{-}$ to $4^{-}$ $\alpha $ strength
function ratio. Trying to reduce to calculated ratio by only the smaller
amount needed to reproduce the measured ratio in the region below $E_{n}=300$
eV quickly leads to strength functions which are orders of magnitude larger
than the observed ones (and hence also to an inferior description of the
cross section). Secondly, it is impossible to obtain a $3^{-}$ to $4^{-}$ $%
\alpha $ strength function ratio smaller than unity with any Saxon-Woods
potential. In consequence, it is impossible to reproduce the small ratio
observed above $E_{n}=300$ eV with any such calculation. Thus, from this
analysis it appears that a $3^{-}$ to $4^{-}$ $\alpha $ strength function
ratio smaller than unity may be an additional indication of a
non-statistical effect in the data.

Finally, the NON-SMOKER calculation predicts that 78\% of the $^{147}$Sm(%
\textit{n},$\alpha $)$^{144}$Nd cross section is given by transitions from 3$%
^{-}$ states, 12\% from 4$^{-}$ states, and 10\% from other states (higher
partial waves). Of the 3$^{-}$ transitions, 67\% directly populate the
ground state of $^{144}$Nd, which cannot be reached from 4$^{-}$ resonances.
New measurements in which the various $\alpha $ groups are resolved would be
very useful for testing these predictions. The relative contributions of the
various transitions also shows why it is more important for cross section
predictions to reproduce the $3^{-}$ TCs than the $3^{-}$ to $4^{-}$\ ratio.
Therefore, potential P2, which has essentially the same $3^{-}$ to $4^{-}$\
ratio as potential P1 but comes much closer to reproducing the 3$^{-}$ TCs,
yields a cross section in better agreement with the measurements.

\section{Conclusion}

Comparing theory to experiment as $\alpha $ strength functions rather than
as $^{147}$Sm(\textit{n},$\alpha $) cross sections avoids confounding
effects due to the neutron and gamma channels, level densities, and width
fluctuation corrections and therefore can reveal more useful information
about possible improvements to theoretical models by isolating effects due
to the $\alpha $+nucleus potential. Furthermore, separating the data into
the two possible \textit{s}-wave spin states may yield even more information
about the $\alpha $+nucleus potential, because $\alpha $ particles from $%
3^{-}$ resonances have (on average) larger energies than those from $4^{-}$
resonances and hence they sample a different region of the $\alpha $+nucleus
potential. Therefore, differences between the measured and calculated $%
\alpha $ strength functions for the two different \textit{s}-wave spin
states should be useful for future improvements in the $\alpha $+nucleus
potential. An improved $\alpha $+nucleus potential would be very useful for
astrophysical applications.

Interestingly, it is clear from the data presented in Fig. \ref%
{Sm147GaGn0AndGgvsE} that the extracted $\alpha $ widths exhibit
fluctuations different from the expected behavior and hint at possible
non-statistical effects. One striking feature is that the $\alpha $ widths
exhibit peaks and/or regions of large fluctuations as a function of neutron
energy instead of the expected random fluctuations intermediate to that for
neutron and $\gamma $ widths. Perhaps these peaks are a manifestation of a
nuclear structure effect in the $^{148}$Sm compound nucleus that is not
observable in decay channels other than the $\alpha $ channel. One important
difference between the decay of the $^{148}$Sm compound nucleus into the $%
\alpha $ channel compared to decay into the neutron or $\gamma $ channels is
the large Coulomb barrier that the $\alpha $ particles must overcome. Hence,
it is possible that the Coulomb barrier for $\alpha $ decay could act as a
lever arm enhancing the signature of nuclear structure effects that are too
subtle to be observed in other channels. For example, $\alpha $ decay could
be enhanced for compound states with significant deformation and for $J^{\pi
}=4^{-}$ states of significant collectivity (because they decay mainly to
the $I^{\pi }=2^{+}$ first excited state of $^{144}$Nd).

Further evidence for possible non-statistical behavior of the $\alpha $
widths is revealed when the resonances are separated according to spin as
shown in Figs. \ref{CumuAlphAndNeutWidFig} through \ref{AvgAlStrFunRat}. The
striking disagreement with theoretical expectations shown in these figures
depends on reliable spin assignments for the resonances. The spin
assignments used in these figures rely on the results of Ref. \cite{Ge93},
the good match between the energy scale for this reference and our work, and
the crude pulse-height information from our experiment for resonances having
large $\alpha $ widths. Below approximately 300 eV, the spin assignments
used herein should be very reliable. Therefore, it is very interesting that
our exploratory calculations show that, even in this energy range, it is not
possible to reproduce the observed $\alpha $ strength functions as well as
the $3^{-}$ to $4^{-}$ $\alpha $ strength ratio at the same time using a
Saxon-Woods potential.

The $\alpha $-width distributions shown in Figs. \ref{CumuAlphAndNeutWidFig}
and \ref{PTDistAlAndNeutFig}, as well as the striking change in the $3^{-}$
to $4^{-}$ $\alpha $ strength ratio near 300 eV shown in Fig. \ref%
{AvgAlStrFunRat} depend on reliable spin assignments above 300 eV. For a
number of reasons, the spin assignments in this region may not be as
reliable, so it would be very useful to make new $^{147}$Sm(\textit{n},$%
\alpha $) measurements with thinner samples to check spin assignments,
especially for those resonances having the largest $\alpha $ widths. In such
measurements, resonances having visible $\alpha _{0}$ groups to the ground
state of $^{144}$Nd unambiguously could be assigned as $J=3$. The $\alpha $%
-particle spectra for the few resonances below $E_{n}=185$ eV that have been
reported \cite{Po72} are in agreement with the accepted \cite{Su98} spin
assignments, except possibly the 58.130-eV resonance (57.9 eV in Ref. \cite%
{Po72}), which is assigned $J^{\pi }=3^{-}$ but has an almost invisible $%
\alpha _{0}$ group. New neutron capture and total cross section measurements
on $^{147}$Sm also could be useful. New measurements with higher resolution
and sensitivity could reduce uncertainties in the $\alpha $ widths by
identifying some of the missing resonances in the relevant energy range and
by providing more $\gamma $ widths as well as more precise neutron widths.
Finally, (\textit{n},$\alpha $), (\textit{n},$\gamma $), and neutron total
cross sections on other nuclides in this mass region should be very useful
in shining more light on this interesting problem.

\begin{acknowledgments}
We would like to thank J. A. Harvey, J. E. Lynn, Yu. P. Popov, S. Raman, and
F.-K. Thielemann for fruitful discussions. This work was supported in part
by the U.S. Department of Energy under Contract No. DE-AC05-00OR22725 with
UT-Battelle, LLC, and by the Swiss NSF (grant 2000-061031.02). T. R.
acknowledges support by a PROFIL professorship from the Swiss NSF (grant
2024-067428.01).
\end{acknowledgments}

\newif\ifabfull\abfulltrue


\begin{thebibliography}{28}
\expandafter\ifx\csname natexlab\endcsname\relax\def\natexlab#1{#1}\fi
\expandafter\ifx\csname bibnamefont\endcsname\relax
  \def\bibnamefont#1{#1}\fi
\expandafter\ifx\csname bibfnamefont\endcsname\relax
  \def\bibfnamefont#1{#1}\fi
\expandafter\ifx\csname citenamefont\endcsname\relax
  \def\citenamefont#1{#1}\fi
\expandafter\ifx\csname url\endcsname\relax
  \def\url#1{\texttt{#1}}\fi
\expandafter\ifx\csname urlprefix\endcsname\relax\def\urlprefix{URL }\fi
\providecommand{\bibinfo}[2]{#2}
\providecommand{\eprint}[2][]{\url{#2}}


\bibitem[Hoffman et~al.(1999)Hoffman, Woosley, Weaver, Rauscher, and
Thielemann]{Ho99} 
\bibinfo{author}{\bibfnamefont{R.~D.}
\bibnamefont{Hoffman}}, 
\bibinfo{author}{\bibfnamefont{S.~E.}
\bibnamefont{Woosley}}, 
\bibinfo{author}{\bibfnamefont{T.~A.}
\bibnamefont{Weaver}}, \bibinfo{author}{\bibfnamefont{T.}~%
\bibnamefont{Rauscher}}, and 
\bibinfo{author}{\bibfnamefont{F.-K.}
\bibnamefont{Thielemann}}, \bibinfo{journal}{Astrophys. J.} \textbf{%
\bibinfo{volume}{521}}, \bibinfo{pages}{735} (\bibinfo{year}{1999}).

\bibitem[Rauscher et~al.(2002)Rauscher, Heger, Hoffman, and Woosley]{Ra2002} %
\bibinfo{author}{\bibfnamefont{T.}~\bibnamefont{Rauscher}}, %
\bibinfo{author}{\bibfnamefont{A.}~\bibnamefont{Heger}}, \bibinfo{author}{%
\bibfnamefont{R.~D.} \bibnamefont{Hoffman}}, and 
\bibinfo{author}{\bibfnamefont{S.~E.}
  \bibnamefont{Woosley}}, \bibinfo{journal}{Astrophys. J.} \textbf{%
\bibinfo{volume}{576}}, \bibinfo{pages}{323} (\bibinfo{year}{2002}).

\bibitem[Gledenov et~al.(2000)Gledenov, Koehler, Andrzejewski, Guber, and
Rauscher]{Gl2000} 
\bibinfo{author}{\bibfnamefont{Y.~M.}
\bibnamefont{Gledenov}}, 
\bibinfo{author}{\bibfnamefont{P.~E.}
\bibnamefont{Koehler}}, \bibinfo{author}{\bibfnamefont{J.}~%
\bibnamefont{Andrzejewski}}, 
\bibinfo{author}{\bibfnamefont{K.~H.}
\bibnamefont{Guber}}, and 
\bibinfo{author}{\bibfnamefont{T.}
\bibnamefont{Rauscher}}, \bibinfo{journal}{Phys. Rev. C} \textbf{%
\bibinfo{volume}{62}}, \bibinfo{pages}{042801(R)} (\bibinfo{year}{2000}).

\bibitem[Popov et~al.(1972)Popov, Przytula, Rumi, Stempinski, and Frontasyeva%
]{Po72} \bibinfo{author}{\bibfnamefont{Y.~P.} \bibnamefont{Popov}}, %
\bibinfo{author}{\bibfnamefont{M.}~\bibnamefont{Przytula}}, %
\bibinfo{author}{\bibfnamefont{R.~F.} \bibnamefont{Rumi}}, %
\bibinfo{author}{\bibfnamefont{M.}~\bibnamefont{Stempinski}}, and %
\bibinfo{author}{\bibfnamefont{M.}~\bibnamefont{Frontasyeva}}, %
\bibinfo{journal}{Nucl. Phys.} \textbf{\bibinfo{volume}{A188}}, %
\bibinfo{pages}{212} (\bibinfo{year}{1972}).

\bibitem[Balabanov et~al.(1976)Balabanov, Gledenov, Chol, Popov, and Semenov]%
{Ba76} \bibinfo{author}{\bibfnamefont{N.~P.} \bibnamefont{Balabanov}}, %
\bibinfo{author}{\bibfnamefont{Y.~M.} \bibnamefont{Gledenov}}, %
\bibinfo{author}{\bibfnamefont{P.~H.} \bibnamefont{Chol}}, %
\bibinfo{author}{\bibfnamefont{Y.~P.} \bibnamefont{Popov}}, and %
\bibinfo{author}{\bibfnamefont{V.~G.} \bibnamefont{Semenov}}, %
\bibinfo{journal}{Nucl. Phys.} \textbf{\bibinfo{volume}{A261}}, %
\bibinfo{pages}{35} (\bibinfo{year}{1976}).

\bibitem[Andzheevski et~al.(1980)Andzheevski, Tkhan', Vtyurin, Koreivo,
Popov, and Stempin'ski]{An80} \bibinfo{author}{\bibfnamefont{Y.}~%
\bibnamefont{Andzheevski}}, 
\bibinfo{author}{\bibfnamefont{V.~K.}
\bibnamefont{Tkhan'}}, 
\bibinfo{author}{\bibfnamefont{V.~A.}
\bibnamefont{Vtyurin}}, \bibinfo{author}{\bibfnamefont{A.}~%
\bibnamefont{Koreivo}}, 
\bibinfo{author}{\bibfnamefont{Y.~P.}
\bibnamefont{Popov}}, and \bibinfo{author}{\bibfnamefont{M.}~%
\bibnamefont{Stempin'ski}}, \bibinfo{journal}{Yad. Fiz.} \textbf{%
\bibinfo{volume}{32}}, \bibinfo{pages}{1496} (\bibinfo{year}{1980}).

\bibitem[Antonov et~al.(1984)Antonov, Gledenov, Marinova, Popov, and Rigol]%
{An84} \bibinfo{author}{\bibfnamefont{A.}~\bibnamefont{Antonov}}, %
\bibinfo{author}{\bibfnamefont{Y.~M.} \bibnamefont{Gledenov}}, %
\bibinfo{author}{\bibfnamefont{S.}~\bibnamefont{Marinova}}, %
\bibinfo{author}{\bibfnamefont{Y.~P.} \bibnamefont{Popov}}, and %
\bibinfo{author}{\bibfnamefont{H.}~\bibnamefont{Rigol}}, %
\bibinfo{journal}{Yad. Fiz} \textbf{\bibinfo{volume}{39}}, %
\bibinfo{pages}{794} (\bibinfo{year}{1984}).

\bibitem[Kvitek and Popov(1970)]{Kv70} \bibinfo{author}{\bibfnamefont{J.}~%
\bibnamefont{Kvitek}} and 
\bibinfo{author}{\bibfnamefont{Y.~P.}
\bibnamefont{Popov}}, \bibinfo{journal}{Nucl. Phys.} \textbf{%
\bibinfo{volume}{A154}}, \bibinfo{pages}{177} (\bibinfo{year}{1970}).

\bibitem[Rapp et~al.(2003)Rapp, Koehler, K{\"a}ppeler, and Raman]{Ra2003} %
\bibinfo{author}{\bibfnamefont{W.}~\bibnamefont{Rapp}}, \bibinfo{author}{%
\bibfnamefont{P.~E.} \bibnamefont{Koehler}}, \bibinfo{author}{%
\bibfnamefont{F.}~\bibnamefont{K{\"a}ppeler}}, and \bibinfo{author}{%
\bibfnamefont{S.}~\bibnamefont{Raman}}, \bibinfo{journal}{Phys. Rev. C} 
\textbf{\bibinfo{volume}{68}}, \bibinfo{pages}{??} (\bibinfo{year}{2003}).

\bibitem[Peelle et~al.(1982)Peelle, Harvey, Maienschein, Weston, Olsen,
Larson, and Macklin]{Pe82} 
\bibinfo{author}{\bibfnamefont{R.~W.}
\bibnamefont{Peelle}}, 
\bibinfo{author}{\bibfnamefont{J.~A.}
\bibnamefont{Harvey}}, 
\bibinfo{author}{\bibfnamefont{F.~C.}
\bibnamefont{Maienschein}}, 
\bibinfo{author}{\bibfnamefont{L.~W.}
\bibnamefont{Weston}}, 
\bibinfo{author}{\bibfnamefont{D.~K.}
\bibnamefont{Olsen}}, 
\bibinfo{author}{\bibfnamefont{D.~C.}
\bibnamefont{Larson}}, and 
\bibinfo{author}{\bibfnamefont{R.~L.}
  \bibnamefont{Macklin}}, \bibinfo{type}{Tech. Rep.} %
\bibinfo{number}{ORNL/TM-8225}, 
\bibinfo{institution}{Oak Ridge National
  Laboratory} (\bibinfo{year}{1982}).

\bibitem[Bockhoff et~al.(1990)Bockhoff, Carlson, Wasson, Harvey, and Larson]%
{Bo90} \bibinfo{author}{\bibfnamefont{K.~H.} \bibnamefont{Bockhoff}}, %
\bibinfo{author}{\bibfnamefont{A.~D.} \bibnamefont{Carlson}}, %
\bibinfo{author}{\bibfnamefont{O.~A.} \bibnamefont{Wasson}}, %
\bibinfo{author}{\bibfnamefont{J.~A.} \bibnamefont{Harvey}}, and 
\bibinfo{author}{\bibfnamefont{D.~C.}
  \bibnamefont{Larson}}, \bibinfo{journal}{Nucl. Sci. and Eng.} \textbf{%
\bibinfo{volume}{106}}, \bibinfo{pages}{192} (\bibinfo{year}{1990}).

\bibitem[Guber et~al.(1997)Guber, Larson, Koehler, Spencer, Raman, Harvey,
Hill, Lewis, and Winters]{Gu97b} 
\bibinfo{author}{\bibfnamefont{K.~H.}
\bibnamefont{Guber}}, 
\bibinfo{author}{\bibfnamefont{D.~C.}
\bibnamefont{Larson}}, 
\bibinfo{author}{\bibfnamefont{P.~E.}
\bibnamefont{Koehler}}, 
\bibinfo{author}{\bibfnamefont{R.~R.}
\bibnamefont{Spencer}}, \bibinfo{author}{\bibfnamefont{S.}~%
\bibnamefont{Raman}}, 
\bibinfo{author}{\bibfnamefont{J.~A.}
\bibnamefont{Harvey}}, 
\bibinfo{author}{\bibfnamefont{N.~W.}
\bibnamefont{Hill}}, 
\bibinfo{author}{\bibfnamefont{T.~A.}
\bibnamefont{Lewis}}, and 
\bibinfo{author}{\bibfnamefont{R.~R.}
\bibnamefont{Winters}}, in \emph{%
\bibinfo{booktitle}{International Conference on Nuclear Data for
  Science and Technology}}, edited by \bibinfo{editor}{\bibfnamefont{G.}~%
\bibnamefont{Reffo}}, \bibinfo{editor}{\bibnamefont{A.Ventura}}, and %
\bibinfo{editor}{\bibfnamefont{C.}~\bibnamefont{Grandi}} (%
\bibinfo{publisher}{Societa Italiana di Fisica}, \bibinfo{address}{Bologna}, %
\bibinfo{year}{1997}), p. \bibinfo{pages}{559}.

\bibitem[Koehler et~al.(1995)Koehler, Harvey, and Hill]{Ko95} %
\bibinfo{author}{\bibfnamefont{P.~E.} \bibnamefont{Koehler}}, %
\bibinfo{author}{\bibfnamefont{J.~A.} \bibnamefont{Harvey}}, and %
\bibinfo{author}{\bibfnamefont{N.~W.} \bibnamefont{Hill}}, %
\bibinfo{journal}{Nucl. Instr. and Meth.} \textbf{\bibinfo{volume}{A361}}, %
\bibinfo{pages}{270} (\bibinfo{year}{1995}).

\bibitem[Carlson et~al.(1993)Carlson, Poenitz, Hale, Peele, Dodder, Fu, and
Mannhart]{Ca93} \bibinfo{author}{\bibfnamefont{A.~D.} \bibnamefont{Carlson}}%
, \bibinfo{author}{\bibfnamefont{W.~P.} \bibnamefont{Poenitz}}, %
\bibinfo{author}{\bibfnamefont{G.~M.} \bibnamefont{Hale}}, %
\bibinfo{author}{\bibfnamefont{R.~W.} \bibnamefont{Peele}}, %
\bibinfo{author}{\bibfnamefont{D.~C.} \bibnamefont{Dodder}}, %
\bibinfo{author}{\bibfnamefont{C.~Y.} \bibnamefont{Fu}}, and %
\bibinfo{author}{\bibfnamefont{W.}~\bibnamefont{Mannhart}}, %
\bibinfo{type}{Tech. Rep.}, 
\bibinfo{institution}{National Institute of
  Standards and Technology Report NISTIR-5177} (\bibinfo{year}{1993}), %
\bibinfo{note}{1993}.

\bibitem[Ziegler and Biersack(1999)]{Zi99} \bibinfo{author}{%
\bibfnamefont{J.~F.} \bibnamefont{Ziegler}} and \bibinfo{author}{%
\bibfnamefont{J.~P.} \bibnamefont{Biersack}}, \emph{%
\bibinfo{title}{{SRIM}
2000}} (\bibinfo{year}{1999}).

\bibitem[Larson(2000)]{La2000} 
\bibinfo{author}{\bibfnamefont{N.~M.}
\bibnamefont{Larson}}, \bibinfo{type}{Tech. Rep.} %
\bibinfo{number}{ORNL/TM-2000/252}, 
\bibinfo{institution}{Oak Ridge National
Laboratory, 2000} (\bibinfo{year}{2000}).

\bibitem[Sukhoruchkin et~al.(1998)Sukhoruchkin, Soroko, and Deriglazov]%
{Su98} \bibinfo{author}{\bibfnamefont{S.~I.} \bibnamefont{Sukhoruchkin}}, %
\bibinfo{author}{\bibfnamefont{Z.~N.} \bibnamefont{Soroko}}, and 
\bibinfo{author}{\bibfnamefont{V.~V.}
  \bibnamefont{Deriglazov}}, \emph{%
\bibinfo{title}{Low Energy Neutron
Physics}} (\bibinfo{publisher}{Springer-Verlag}, \bibinfo{address}{Berlin}, %
\bibinfo{year}{1998}).

\bibitem[Georgiev et~al.(1993)Georgiev, Zamyatnin, Pikelner, Muradian,
Grigoriev, Madjarski, and Janeva]{Ge93} \bibinfo{author}{\bibfnamefont{G.}~%
\bibnamefont{Georgiev}}, 
\bibinfo{author}{\bibfnamefont{Y.~S.}
\bibnamefont{Zamyatnin}}, 
\bibinfo{author}{\bibfnamefont{L.~B.}
\bibnamefont{Pikelner}}, 
\bibinfo{author}{\bibfnamefont{G.~V.}
\bibnamefont{Muradian}}, 
\bibinfo{author}{\bibfnamefont{Y.~V.}
\bibnamefont{Grigoriev}}, \bibinfo{author}{\bibfnamefont{T.}~%
\bibnamefont{Madjarski}}, and \bibinfo{author}{\bibfnamefont{N.}~%
\bibnamefont{Janeva}}, \bibinfo{journal}{Nucl. Phys.} \textbf{%
\bibinfo{volume}{A565}}, \bibinfo{pages}{643} (\bibinfo{year}{1993}).

\bibitem[J.~W.~Codding et~al.(1971)J.~W.~Codding, Tromp, and Simpson]{Co71} %
\bibinfo{author}{\bibfnamefont{J.}~\bibnamefont{J.~W.~Codding}}, %
\bibinfo{author}{\bibfnamefont{R.~L.} \bibnamefont{Tromp}}, and %
\bibinfo{author}{\bibfnamefont{F.~B.} \bibnamefont{Simpson}}, %
\bibinfo{journal}{Nucl. Sci. Eng.} \textbf{\bibinfo{volume}{43}}, %
\bibinfo{pages}{58} (\bibinfo{year}{1971}).

\bibitem[Eiland et~al.(1974)Eiland, Weinstein, and Seeman]{Ei74} %
\bibinfo{author}{\bibfnamefont{H.~M.} \bibnamefont{Eiland}}, %
\bibinfo{author}{\bibfnamefont{S.}~\bibnamefont{Weinstein}}, and 
\bibinfo{author}{\bibfnamefont{K.~W.}
  \bibnamefont{Seeman}}, \bibinfo{journal}{Nucl. Sci. Eng.} \textbf{%
\bibinfo{volume}{54}}, \bibinfo{pages}{286} (\bibinfo{year}{1974}).

\bibitem[Mizumoto(1981)]{Mi81} \bibinfo{author}{\bibfnamefont{M.}~%
\bibnamefont{Mizumoto}}, \bibinfo{journal}{Nucl. Phys.} \textbf{%
\bibinfo{volume}{A357}}, \bibinfo{pages}{90} (\bibinfo{year}{1981}).

\bibitem[Mughabghab(1984)]{Mu84} 
\bibinfo{author}{\bibfnamefont{S.~F.}
\bibnamefont{Mughabghab}}, \emph{\bibinfo{title}{Neutron Cross Sections}} (%
\bibinfo{publisher}{Academic
  Press}, \bibinfo{address}{New York}, \bibinfo{year}{1984}).

\bibitem[Mughabghab et~al.(1981)Mughabghab, Divadeenam, and Holden]{Mu81} %
\bibinfo{author}{\bibfnamefont{S.~F.} \bibnamefont{Mughabghab}}, %
\bibinfo{author}{\bibfnamefont{M.}~\bibnamefont{Divadeenam}}, and 
\bibinfo{author}{\bibfnamefont{N.~E.}
  \bibnamefont{Holden}}, \emph{\bibinfo{title}{Neutron Cross Sections}},
vol.~\bibinfo{volume}{1} (\bibinfo{publisher}{Academic}, %
\bibinfo{address}{New York}, \bibinfo{year}{1981}).

\bibitem[Porter and Thomas(1956)]{Po56} \bibinfo{author}{%
\bibfnamefont{C.~E.} \bibnamefont{Porter}} and \bibinfo{author}{%
\bibfnamefont{R.~G.} \bibnamefont{Thomas}}, \bibinfo{journal}{Phys. Rev.} 
\textbf{\bibinfo{volume}{104}}, \bibinfo{pages}{483} (\bibinfo{year}{1956}).

\bibitem[Gledenov et~al.(2002)Gledenov, Koehler, Andrzejewski, Popov, and
Gledenov]{Gl2002} 
\bibinfo{author}{\bibfnamefont{Y.~M.}
\bibnamefont{Gledenov}}, 
\bibinfo{author}{\bibfnamefont{P.~E.}
\bibnamefont{Koehler}}, \bibinfo{author}{\bibfnamefont{J.}~%
\bibnamefont{Andrzejewski}}, 
\bibinfo{author}{\bibfnamefont{Y.~P.}
\bibnamefont{Popov}}, and 
\bibinfo{author}{\bibfnamefont{R.~Y.}
\bibnamefont{Gledenov}}, 
\bibinfo{journal}{Nucl. Sci. and Tech., Supplement
2} \textbf{\bibinfo{volume}{1}}, \bibinfo{pages}{358} (\bibinfo{year}{2002}).

\bibitem[Koehler et~al.(2001)Koehler, Gledenov, Andrzejewski, Guber, Raman,
and Rauscher]{Ko2001} 
\bibinfo{author}{\bibfnamefont{P.~E.}
\bibnamefont{Koehler}}, 
\bibinfo{author}{\bibfnamefont{Y.~M.}
\bibnamefont{Gledenov}}, \bibinfo{author}{\bibfnamefont{J.}~%
\bibnamefont{Andrzejewski}}, 
\bibinfo{author}{\bibfnamefont{K.~H.}
\bibnamefont{Guber}}, \bibinfo{author}{\bibfnamefont{S.}~\bibnamefont{Raman}}%
, and \bibinfo{author}{\bibfnamefont{T.}~\bibnamefont{Rauscher}}, %
\bibinfo{journal}{Nucl. Phys.} \textbf{\bibinfo{volume}{A688}}, %
\bibinfo{pages}{86c} (\bibinfo{year}{2001}).

\bibitem[Popov et~al.(1970)Popov, Prztula, Rumi, Stempinski, Florek, and
Furman]{Po70} \bibinfo{author}{\bibfnamefont{Y.~P.} \bibnamefont{Popov}}, %
\bibinfo{author}{\bibfnamefont{M.}~\bibnamefont{Prztula}}, %
\bibinfo{author}{\bibfnamefont{R.~F.} \bibnamefont{Rumi}}, %
\bibinfo{author}{\bibfnamefont{M.}~\bibnamefont{Stempinski}}, %
\bibinfo{author}{\bibfnamefont{M.}~\bibnamefont{Florek}}, and %
\bibinfo{author}{\bibfnamefont{V.~I.} \bibnamefont{Furman}}, in \emph{%
\bibinfo{booktitle}{Nuclear Data for Reactors}} (%
\bibinfo{publisher}{International Atomic Energy Agency}, %
\bibinfo{address}{Vienna}, \bibinfo{year}{1970}), p. \bibinfo{pages}{669}.

\bibitem{hauser} W. Hauser and H. Feshbach, Phys.\ Rev. \textbf{87}, 366
(1952).

\bibitem{gad92} E. Gadioli and P. E. Hodgson, \textit{Pre-Equilibrium
Nuclear Reactions} (Clarendon Press, Oxford, 1992).

\bibitem{rau98} T. Rauscher and F.-K. Thielemann, in \textit{Stellar
Evolution, Stellar Explosions, and Galactic Chemical Evolution}, edited by\
A. Mezzacappa (IOP, Bristol, 1998), p.\ 519.

\bibitem{rau00} T. Rauscher and F.-K. Thielemann, Atomic Data Nucl.\ Data
Tables \textbf{75}, 1 (2000).

\bibitem{rau01} T. Rauscher and F.-K. Thielemann, Atomic Data Nucl.\ Data
Tables \textbf{79}, 47 (2001).

\bibitem{McF} L. McFadden and G. R. Satchler, Nucl.\ Phys. \textbf{84}, 177
(1966).

\bibitem{Fro02} C. Fr\"ohlich, diploma thesis, University of Basel,
Switzerland (unpublished).

\bibitem{raufro03} T. Rauscher, C. Fr\"{o}hlich, K. H. Guber, in \textit{%
Capture Gamma-Ray Spectroscopy and Related Topics}, edited by J. Kvasil, P.
Cejnar, and M. Krticka (World Scientific, Singapore, 2003), p. 781;
nucl-th/0302046.

\bibitem[Rauscher(2003)]{Ra2003a} \bibinfo{author}{\bibfnamefont{T.}~%
\bibnamefont{Rauscher}}, \bibinfo{journal}{Nucl. Phys.} \textbf{%
\bibinfo{volume}{A719}}, \bibinfo{pages}{73c} (\bibinfo{year}{2003}); Nucl.\
Phys.\ \textbf{A725}, 295 (2003) (Erratum).
\end{thebibliography}
\end{document}